\documentclass[11pt]{article}%%%%where rsproca is the template name
\usepackage{xcolor}
\usepackage{ulem}
\usepackage{dcolumn}
\usepackage{bm}
\usepackage{amsfonts}
\usepackage{amsbsy}
\usepackage{graphicx}
\usepackage{psfrag}
\usepackage{verbatim}
\usepackage{color}
\usepackage{amsmath}
\usepackage{epsfig,bm,epsf,float}
\usepackage{mathrsfs}
\usepackage{latexsym}
\usepackage{mathrsfs}
\usepackage{soul}
\usepackage{cancel}
\usepackage{newtxtext}
\usepackage{newtxmath}
\usepackage[numbers]{natbib}
\usepackage{hyperref}
\usepackage{authblk}
\usepackage{blindtext}
\usepackage[a4paper, total={5.3in, 9.8in}]{geometry}

\usepackage[stdsubgroups]{nomencl} % Nomenclature
\makenomenclature

\usepackage{ifthen}
\renewcommand{\nomgroup}[1]{%
\ifthenelse{\equal{#1}{G}}{\item[\textbf{Greek Symbols}]}{%
\ifthenelse{\equal{#1}{S}}{\item[\textbf{Subscripts}]}{}}}

\graphicspath{{./fig/}}
\newcommand{\ms}{\kern.10em\relax}
\usepackage{bm}
\usepackage{mathrsfs}

\def\bea{\begin{eqnarray}}
\def\eea{\end{eqnarray}}
\def\th{\theta}

\def\th{\theta}
\def\ee{\mathrm{e}}

\def\bea{\begin{eqnarray}}
\def\eea{\end{eqnarray}}

\def\eps{\epsilon}

\def\H{\mathcal{H}}

\def\sig{\sigma}

\def\cb{\color{black}}

\begin{document}

%%%% Article title to be placed here
\title{The Moffatt-Pukhnachev flow: a new twist on an old problem. }

\author[1]{Antonio J. Bárcenas-Luque}
\author[2]{Mark G. Blyth}

\affil[1]{Department of Mechanics of Structures and Hydraulic Engineering, University of Granada, 18001 Granada, Spain.}
\affil[2]{School of Mathematics, University of East Anglia, Norwich, England, NR4 7TJ.}

\maketitle

%%%% Abstract text to be placed here %%%%%%%%%%%%
\begin{abstract}
The flow of a thin viscous film on the outside of a horizontal circular cylinder, whose angular 
velocity consists of a steady part and an oscillatory part, is investigated. In the absence of the oscillatory
component the problem reduces to that 
originally studied by Moffatt \cite{moffatt1977behaviour} and Pukhnachev \cite{pukhnachev1977motion}.
Surface tension is neglected. The characteristic equations for the thin-film governing equation are solved 
numerically for a range of values of the oscillation amplitude and 
frequency. A blow-up map charted in amplitude-frequency space for a particular characteristic 
reveals highly intricate fractal-like structures exhibiting self-similarity. In general, starting from a specified 
initial profile the surface of the film reaches a slope singularity at a finite time and tends to overturn provoking 
the formation of a shock. It is shown that the overturning time can be considerably delayed by a careful 
choice of the starting profile. The high-frequency and low-frequency limits are examined asymptotically using 
a multiple-scales approach. At high frequency the analysis suggests that an appropriate choice of initial 
profile can substantially delay the overturning time and even yield a time-periodic solution that does not 
overturn. In the low-frequency limit it is possible to construct a quasi-periodic solution that does not overturn if 
the oscillation amplitude lies below a threshold value. Above this value the solution tends inexorably toward 
blow-up. It is shown how solutions exhibiting either a single-shock or a double-shock may be constructed in common with the steadily rotating cylinder problem.
\end{abstract}
%%%%%%%%%%%%%%%%%%%%%%%%%%%

%\rsbreak

%%%%%%%%%% Insert the texts which can accomdate on firstpage in the tag "fmtext" %%%%%

\section{Introduction}
Moffat \cite{moffatt1977behaviour} studied the dynamics of a thin viscous liquid film 
coating the outside of a horizontal circular cylinder that is rotating about its axis at a constant rate.
This problem has important practical applications ranging from industrial processes (Ribatski \& Jacobi \cite{Ribatski2005_refrigeration}) to applications in the art world (Herczynski {\it et al.} \cite{pollock2011}).
In the absence of rotation the film will eventually drip under the action of gravity. Moffatt showed that rotation 
prevents dripping if the angular velocity of the cylinder exceeds a threshold value that depends on the 
kinematic viscosity of the liquid, the radius of the cylinder, and the acceleration due to gravity. This is in line 
with everyday experience: we can prevent honey from dripping off a spoon by rotating the spoon. However, it 
is tricky to rotate the spoon continuously at a constant rate, and in practice one tends 
rather to twist it back and forth, endowing the spoon with a time-dependent angular velocity.

%This study provided key insights into the balance between centrifugal forces, gravity, surface tension, and viscosity, laying the groundwork for subsequent investigations into related fluid dynamics problems.

Working on the basis of lubrication theory, Moffatt \cite{moffatt1977behaviour} derived a nonlinear model 
equation for the film thickness and showed that it has a steady solution if the aforementioned threshold 
criterion is met. Working around the same time, Pukhnachev \cite{pukhnachev1977motion} derived a more 
general version of the governing equation that incorporated the effect of surface tension, and  
also demonstrated the existence and uniqueness of a steady solution. The steady solution describes a film 
profile which is stationary in the laboratory frame. Assuming counterclockwise rotation, the film exhibits a 
bulge in thickness on the right side of the cylinder and is thinner on the left side (see 
figure~\ref{fig:sketch}). Moffatt \cite{moffatt1977behaviour} also described the results of experiments that 
revealed the importance of transverse instability, which manifests as a sequence of liquid lobes spaced out 
along the axis of the cylinder. 

Numerous papers have followed examining various aspects of a problem that has proved to be very rich from a dynamics perspective. Hinch and Kelmanson \cite{Hinch_Kelmanson_2003} used asymptotic methods to show that 
surface perturbations decay and drift over a four time-scale cascade. Hinch, Kelmanson \& Metcalfe 
\cite{hinch2004shock} probed these results further, focusing on shock formation in the zero surface-tension 
case and providing an estimate of the shock formation time. Hansen \& Kelmanson \cite{hansen1994steady} 
used a boundary integral formulation to compute surface profiles under conditions of Stokes flow, allowing for 
films of arbitrary thickness and including surface tension.  
Peterson et al. \cite{Peterson_Jimack_Kelmanson_2001} carried out a comprehensive linear stability 
analysis and revealed the parameter regimes in which steady-state solutions are stable. Duffy and Wilson 
\cite{Duffy_Wilson_1999_curtain} analysed both attached films and curtain flows (for which a film 
falls onto the cylinder from above, curves around, and falls off the bottom),  developed analytical 
approximations, and identified critical flow transitions. Wray and Cimpeanu \cite{wray_cimpeanu_2020} used 
reduced-order modelling techniques to effectively capture the key features of the dynamics when inertia is 
included. Karabut \cite{Karabut_2007} explored two distinct flow regimes depending on the angular velocity, 
which contributed to the classification of flow behaviours.

Evans, Schwartz and Roy \cite{Evans_2004_2dcoating, 
Evans_2005_3dcoating} presented both two- and three-dimensional models for coating flow on a rotating 
cylinder. They examined unsteady behaviours and instabilities.  
Gorla \cite{gorla2000_rupture} focused on the rupture dynamics of non-Newtonian (power-law) films, 
addressing industrially relevant coating processes where film integrity is critical.  
Lopes {\it et al.}  \cite{lopes2018multiple} introduced a new model equation which was derived using a gradient method based on Onsager's variational principle and which includes the full expression for the surface curvature in the capillary stress term. These authors also compared the model predictions for steady flow with full computations of the Stokes equations.
Kelmanson \cite{Kelmanson_2009} extended the Moffatt–Pukhnachev model by including inertial effects, thereby 
increasing its relevance to moderate Reynolds number regimes. 

Weak solutions exhibiting shocks were first 
introduced for the steady Moffatt-Pukhnachev flow by Johnson \cite{johnson1988steady}; see also 
Badali {\it et al.} \cite{Badali2011_shocks_coating} and Benilov {\it et al.} \cite{Benilov_shocks_rimming}.
Such solutions allow the cylinder to support a greater liquid volume than the smooth Moffatt solutions 
\cite{moffatt1977behaviour}. The stability of these weak solutions was studied by O'Brien \cite{o2002linear} 
and Villegas-D\'iaz at al. \cite{villegas2003stability}, who demonstrated that stable configurations occur only 
when the shock is located in the fourth quadrant of the plane. 

The problem of rimming flow, in which the liquid film coats the inside of the cylinder, is also of interest. As 
highlighted by Lopes \cite{lopes2018dynamics} the thin-film equations for rimming flow and for the exterior 
flow problem are identical up to a certain order of approximation in the film thickness parameter, although key differences appear at higher order.
Johnson \cite{johnson1988steady} presented an analysis of steady-state coating flows inside rotating 
horizontal cylinders. O'Brien and Gath \cite{o1998location} 
identified the formation and position of shocks in rimming flows, addressing the occurrence of sharp 
transitions in film thickness. O'Brien \cite{o2002linear} further contributed a linear stability analysis of rimming 
flows, describing conditions under which small disturbances may be amplified. 
%Benilov, O'Brien, and Sazonov \cite{benilov2003new} described the new phenomenon of `explosive 
% instabilty'. 
Villegas-Diaz, Power and Riley \cite{villegas2003stability,villegas2005shocks} examined the stability of 
rimming flows to two-dimensional perturbations, combining analytical and numerical techniques and 
exploring the impact of surface shear on flow stability.

The present paper is devoted to the study of a viscous liquid film that coats the outside of a horizontal 
cylinder which rotates at a constant rate onto which is superimposed oscillations of a certain amplitude and 
frequency. As in the related works discussed above, we herein
assume that the thickness of the liquid film is everywhere much smaller than the radius of the cylinder and 
employ lubrication theory to derive a generalisation of the Moffatt-Pukhnachev equation, which incorporates a 
time-dependent modulation to the rotation rate. Although we discuss the subsequent dynamics in the context 
of the film coating the exterior of the cylinder, the aforementioned equivalence (up to some order in the film 
thickness parameter) between this and the rimming flow problem means that our observations carry over to 
the latter problem to some extent.

The paper is organised as follows. In section \ref{sec:2} we derive the thin-film equation that forms the basis of our model. In section \ref{sec:steady}, we briefly review the  
Moffatt-Puckhnachev flow for constant rotation, taking a dynamical system perspective. 
In section \ref{sec:dynamical_system}, the dynamical system formed from the characteristic equations of our model equation is studied and discussed. In section 
\ref{sec:MPF_modulation} we present solutions to the model equation and study the asymptotic limits of high frequency and low frequency oscillations. Finally, in section \ref{sec:conclusion} we summarise our results.

%%%%%%%%%%%%%%%%%%%%%%%%%%%%%%%%%%%%%

\section{Problem statement}\label{sec:2}

\begin{figure}[!t]
\centering\includegraphics[width=2.35in]{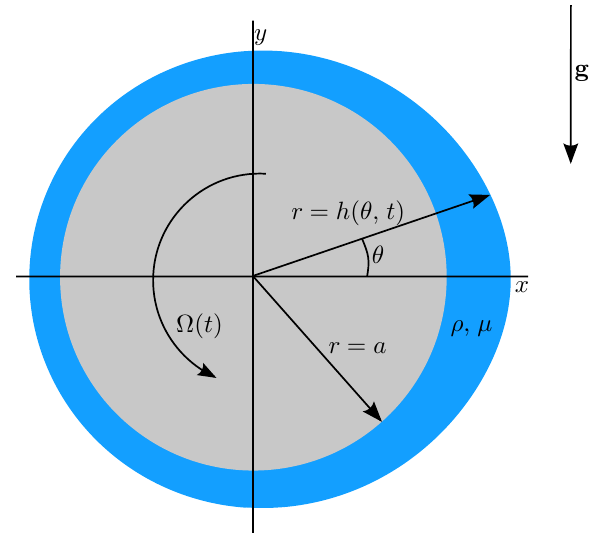}
\caption{Sketch of the flow configuration: a thin viscous liquid film coats the outside of a circular cylinder of radius $a$, which is rotating with angular velocity $\Omega(t)$, where $t$ is time. Gravity acts in the downward vertical direction as shown. The film surface is described in polar coordinates as $r=h(\theta,t)$.}
\label{fig:sketch}
\end{figure}

We consider the flow of a liquid film of viscosity $\mu$ and density $\rho$ 
that coats the exterior of a 
circular cylinder of radius $a$, as is illustrated in 
figure~\ref{fig:sketch}. The motion in the liquid is driven by the 
downwards force of gravity, which acts in the negative $y$ direction, 
and by torsional rotations of the cylinder whose angular velocity is a prescribed periodic function of time, $t$. The flow is assumed to be 
two-dimensional in the $xy$-plane of the cylinder cross-section. The dynamics are described with reference to plane polar coordinates $(r,\theta)$ centred at the cylinder axis with $\theta=0$ aligned with the horizontal. Using thin-film theory Pukhnachev \cite{pukhnachev1977motion} and Moffatt \cite{moffatt1977behaviour} derived an equation for the film thickness when the cylinder rotates at a constant speed. 
Our first goal is to derive a modified version of this equation which accounts for a general angular velocity of the cylinder.
%%%%%%%%%%
%%%%%%%%%%%

%under the lubrication approximation the film thickness $h(\theta,t)$ satisfies the governing equation (see Appendix A)
We define the dimensionless thin-film 
coordinate, $\hat y$, which measures radial 
distance from the surface of the cylinder, such that, with reference to figure~\ref{fig:sketch},
\bea \label{rdefinition}
r = a + \eps a G^{-1/2}\hat y.
\eea
Here we have introduced the dimensionless group $G=\rho g a/\mu\Omega_*$, where 
$g$ is the acceleration due to gravity and $\Omega_*$ is a reference angular 
velocity, and the film thickness parameter 
$\epsilon=\bar h/a$, where $\bar h$ is the mean film thickness. Henceforth we shall assume that $\epsilon \ll 1$ so that the average film thickness $\bar h$ is 
small in comparison with the cylinder radius $a$. 

The velocity $(u,v)$ in the $(r,\theta)$ directions, the pressure $p$, and time $t$ are 
non-dimensionalised by making the replacements
%\bea \label{nondim}
%u = \epsilon^3 G^{-1/2}a\Omega_* \hat u, \qquad
%v = \epsilon^2a\Omega_*\hat v, \qquad p = \rho g a \hat p, \qquad 
%t = \eps^{-2}\Omega_*^{-1}\hat t,
%\eea
\bea \label{nondim}
u \mapsto \epsilon^3 G^{-1/2}a\Omega_* u, \qquad
v \mapsto \epsilon^2a\Omega_*\hat v, \qquad p \mapsto \rho g a p, \qquad 
t \mapsto \eps^{-2}\Omega_*^{-1} t.
\eea
To leading order in $\epsilon$ 
the dimensionless Navier-Stokes equations are
% \begin{align}
% \begin{split}
% 0 =& -\eps^{-1}p_y + \rho a g\cos \theta, \\[0.1in]
% 0 =& -p_\theta - \rho a g\sin \theta + \eps^{-2}(\mu/a)v_{yy},  \\[0.1in] 
% 0 = & u_y + v_\theta,
% \end{split}
% \end{align}
\bea \label{NSred}
0 = p_y, \qquad
%0 = -p_\theta - \rho a g\cos \theta + \eps^{-2}(\mu/a)\, v_{yy},  \qquad
0 = -p_\theta - \cos \theta + v_{yy},  \qquad
0 =  u_y + v_\theta.
\eea
The inertia terms have been neglected in these equations. This is justified provided that $\epsilon^2 Re\ll G$, where the Reynolds number $Re=\rho \Omega_* \bar h^2/\mu$.

At the film surface, $y=h(\theta,t)$, we impose
the kinematic condition,
\bea \label{kincond}
h_{t} + v h_\theta - u = 0,
\eea
as well as the normal and tangential stress conditions
\bea \label{dyncond}
p = 0, \qquad  %-\frac{\epsilon}{G^{3/2}C}\kappa, \qquad
v_y = 0,
\eea
respectively.
The pressure in the air outside of the film has been taken to be zero. Moreover the 
contributions of both the viscous normal stress and the capillary stress in the first condition in \eqref{dyncond} have been 
neglected. These assumptions are justified provided that
$\epsilon^2G^{-1}\ll 1$ and $\eps G^{-3/2}C^{-1}\ll 1$, respectively, where
$C = \mu \Omega_* a/\gamma$ is the capillary number with $\gamma$ the 
coefficient of surface tension. 
The boundary condition on the cylinder, $y=0$, is
\bea \label{surfcond}
%u=0, \qquad v = a \omega \Omega,
u=0, \qquad v = \Omega(t),
\eea
where $\Omega(t)$ is assumed to take the form
\bea \label{omdefinition}
\Omega(t) = 1 + b\cos \sigma t
\eea
for given constants $b>0$ and $\sigma>0$. If $b=0$ then 
the cylinder is rotating at a constant rate; this is the case originally studied by Moffatt and Pukhnachev and we shall henceforth refer to it as the MP problem.

It would appear, then, that there are two relevant time scales in the problem: first, there is the time taken for a 
fluid particle to complete one circuit of the cylinder under steady rotation, i.e. for the MP problem which has a steady flow solution (see  
\cite{moffatt1977behaviour}), and, second, there is the time scale associated with the 
modulational frequency $\sigma$. 
%\cbl {\it Alternatively, for a cylinder performing purely oscillatory oscillations with no mean component, 
%we associate $\omega$ with the angular frequency of the oscillations, and take 
%$\Omega(t) = b\cos t$ for some constant $b$.}\cb

Integrating the governing equations \eqref{NSred}, applying the boundary conditions 
\eqref{dyncond}-\eqref{surfcond}, and inserting the resulting expressions into the kinematic 
condition \eqref{kincond}, we obtain the 
evolution equation
\bea \label{maineq}
h_{t} + Q_\th = 0,
\eea
where
\bea\label{qdefinition}
Q = \int_0^h v\,dy = \Omega(t) h - \frac{1}{3}h^3\cos \th
\eea
is the dimensionless flux in the film. The initial condition is
\bea \label{ic}
h(\theta,0) = \textup{h}_0(\theta)
\eea
for some appropriate choice of the function $\textup{h}_0$. 
%On setting $\Omega=1$ the equation \eqref{maineq} reduces to that derived by Pukhnachev \cite{pukhnachev1977motion} and Moffatt \cite{moffatt1977behaviour}.
%(see also Kelmanson \cite{Kelmanson_2009}, equation 2.5). 

%\cbl {\it It will be helpful to record the velocity components as follows:
%\bea\label{uformula}
%u = \frac{1}{6}\zeta^2(\zeta-3)h^3\sin\th  - \frac{1}{2}\zeta^2 h^2h_\theta\cos \theta,
%\eea
%(\crd MGB: I think the sign on the second term in $u$ should be $+$.\cbl )
%and
%\bea \label{vformula}
%v = \Omega(t) + \frac{1}{2} \zeta(\zeta-2)h^2\cos\th = \Omega(t) \left( 1-3\zeta+\frac{3}{2}\zeta^2\right) + \frac{3Q}{2h}\zeta (2 - \zeta),
%\eea
%where $\zeta = y/h$.
%It is useful to note that the first bracketed term on the far right of \eqref{vformula} has zeros at $\zeta = 1\pm 1/\sqrt{3}\approx 1.58$ and $0.42$, while the second bracketed term is strictly positive (for $\zeta \in [0,1]$).}\cb

In summary we aim to solve \eqref{maineq} with initial condition \eqref{ic} to determine the 
surface profile $h(\theta,t)$ for different choices of the 
parameters $b$ and $\sigma$. Before doing this, however, it is instructive to recall the salient 
details of the steady flow for the MP problem. We do this in the next section 
from a dynamical systems perspective which provides a novel and intuitive way of visualising the solution space.

%Of particular interest is to examine how the steady solution space discussed by Moffatt and Pukhnachev is affected by the 
%time-dependent modulation of the cylinder's angular velocity.

%%%%%%%%%%%%%%%%%%%%%%%%%%%%%%%%%%%%%

\section{The steady Moffatt-Pukhnachev flow}\label{sec:steady}

The steady MP problem is recovered by setting $b=0$ so that the cylinder is rotating at a constant rate. Although it is strictly equal to unity in this case, we find it convenient to retain $\Omega$ in the relevant equations to facilitate later discussion.
The governing equation \eqref{maineq} is 
\bea \label{moffatt}
h_t + \left( \Omega h - \frac{1}{3} h^3  \cos \th \right)_\theta = 0,
%h_t + \left( h - \frac{1}{3} h^3  \cos \th \right)_\theta = 0,
\eea
with initial condition given by \eqref{ic}.
%(Note that, even though it is equal to unity, we retain $\Omega$ in \eqref{moffatt} 
%for later convenience.)
Moffatt \cite{moffatt1977behaviour} showed that, if a certain criterion is met, there exists a steady solution describing a fully attached film. Integrating the steady version of \eqref{moffatt} 
with respect to $\theta$, we obtain
\bea \label{moff2}
\Omega h - \frac{1}{3}h^3 \cos \theta = Q,
%h - \frac{1}{3}h^3 \cos \theta = Q,
\eea
where $Q$ coincides with its definition in \eqref{qdefinition} and is herein constant. 
%
%%%%%%%%%%
\begin{figure}[!t]
\centering\includegraphics[width=5.0in]{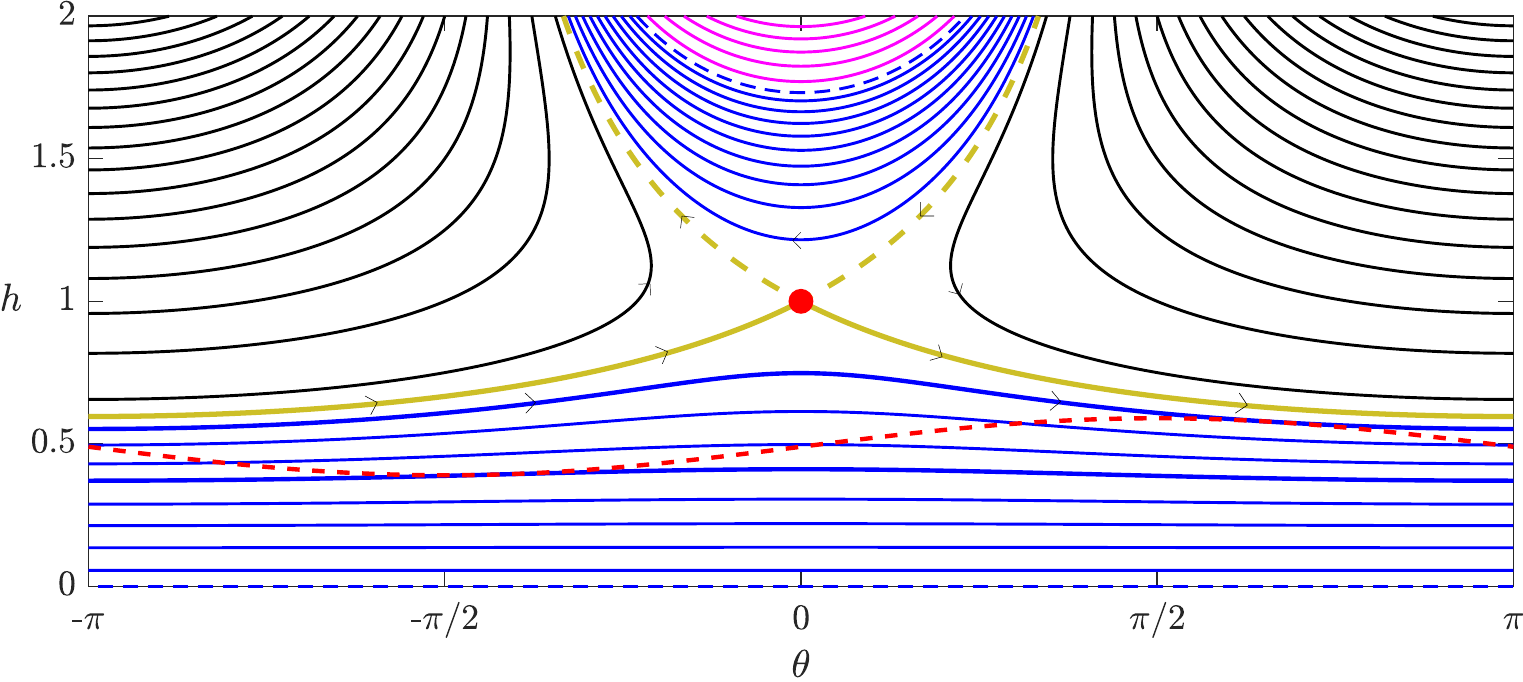}
\caption{One period, $\theta\in [-\pi,\pi)$, of the phase portrait for the system \eqref{chareqs} 
corresponding to the steady MP problem with $\Omega = 1$ $(b=0$). Each 
trajectory is described by \eqref{moff2} and corresponds to a particular value of $Q$. Black 
lines show $Q>2/3$, blue lines show $0<Q<2/3$, and magenta lines show $Q<0$. The blue 
dashed lines indicate $Q=0$. The separatrix, shown with a thick gold line (both solid and 
dashed), corresponds to $Q=Q^*=2/3$; the solid part delineates the boundary between 
regular solutions to the 
characteristic equations \eqref{chareqs} and those exhibiting finite-time blow-up. The 
minimum value of $h$ 
along the separatrix occurs at $\theta=\pm\pi$ and is $h_{\mathrm{min}}=p^{1/3} - 1/
p^{1/3}\approx 0.5961$, where $p=1+\sqrt{2}$. The red dots indicate the location of the 
saddle points for \eqref{chareqs}, and the arrows on the trajectories indicate the direction of 
travel as $\tau$ increases. The broken red line indicates a typical initial film profile, 
$H(\theta)$, bounded by envelope trajectories shown with thick blue lines.}
\label{fig:pplane}
\end{figure}
%%%%%%%%%%%
%
The constant $Q$ level curves for \eqref{moff2} are shown in the phase portrait in 
figure~\ref{fig:pplane}. Critical to note in this figure is the separatrix curve,
shown with a solid gold line, which divides level curves that correspond to physical, fully-
attached solutions (those below the solid gold curve) from those that correspond to unphysical 
solutions (those above it) for which $h$ blows up; in the latter 
case the blow-up has the local form $h \sim a|\theta_c-\theta|^{-1/2}$ for constant $a$, where $\theta_c=-\pi/2$ or $\pi/2$. Solutions with shocks, which include sections of level 
curves above and below the solid separatrix, are also possible and have been discussed by 
previous workers (e.g. Johnson\cite{johnson1988steady}, 
O'Brien \& Gath \cite{o1998location}). We will touch upon these later for the case $b>0$.

\begin{figure}[!t]
\centering\includegraphics[width=5.1in]{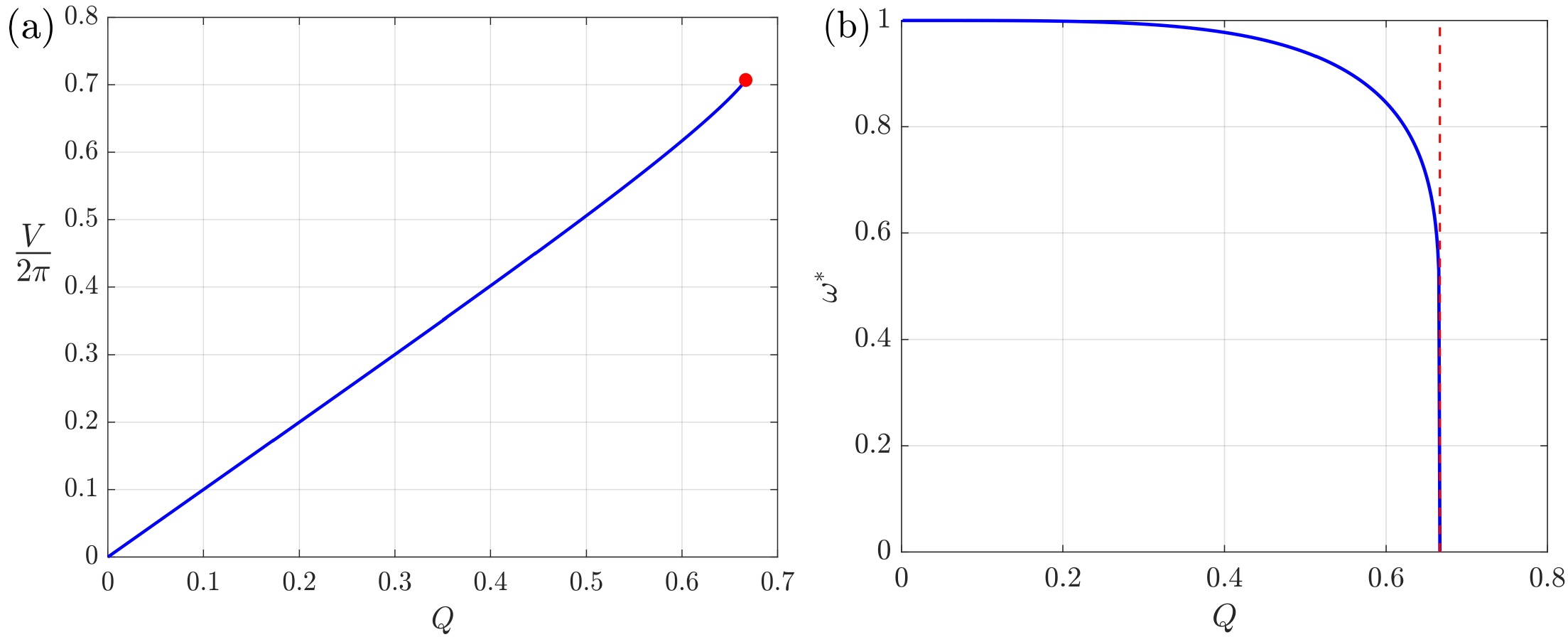}
\caption{(a) The scaled fluid volume, $V/(2\pi)$, plotted against $Q$. The red dot 
indicates the limiting steady volume that obtains at the separatrix in 
Figure~\ref{fig:pplane} when $Q=2/3$. (b) The dependence $\omega_*(Q)$ for the 
MP problem \eqref{moffatt}. The base frequency, $\omega^*$, is related to the period 
$P=2\pi/\omega^*$, being the time taken for a point $(\theta(\tau), h(\tau))$ obeying 
\eqref{chareqs} to complete one $2\pi$-period in $\theta$ of a fixed-$Q$ trajectory in 
the phase plane shown in Figure~\ref{fig:pplane}. }
\label{fig:omega_star}
\end{figure}

A parametric description of the level curves is obtained by solving
the characteristic equations 
for equation \eqref{moffatt}, namely
\bea \label{chareqs}
\frac{d h}{d\tau} = -\frac{1}{3}h^3\sin \theta, \qquad
\frac{d \theta}{d\tau} = \Omega - h^2\cos \theta \equiv M(\theta), \qquad
\frac{dt}{d\tau} = 1,
\eea 
where the independent variable $\tau$ varies continuously along a characteristic. It is 
straightforward to check that \eqref{moff2} is a first integral of \eqref{chareqs}.  We define the period $P(Q)$ of a steady orbit to be the time taken for $\theta$ to change by $2\pi$ radians. By integrating the second equation in \eqref{chareqs}, we find
\bea\label{period}
P(Q) = \int_0^{2\pi} \frac{1}{\Omega-h^2\cos \theta}\,d\theta.
\eea
This fixes a base frequency associated with the steady flow, given by $\omega^\ast(Q)=2\pi/P(Q)$. The dependence of $\omega^\ast$ on $Q$ is graphed in figure \ref{fig:omega_star}(b) where it can be seen that $\omega^\ast\in[0,1]$ and it is monotone decreasing in $Q$. We note the limits $\omega^\ast \to 1$ ($P\to 2\pi$) as $Q\to 0$, and $\omega^\ast\to 0$ ($P\rightarrow\infty$) as $Q\to Q^*$, that is as the separatrix in Figure~\ref{fig:pplane} is approached.

Viewed as a two-dimensional dynamical system
\eqref{chareqs} has a saddle point at $(h,\theta) = (1,\, 0)$. Its stable 
and unstable manifolds are the level curves of \eqref{moff2} with $Q = Q^*$, where
\bea
Q^* = \frac{2}{3}\Omega^{3/2}.
\eea
They are shown with solid gold and dashed gold lines in figure~\ref{fig:pplane}.
An initial condition for 
\eqref{chareqs} at $\tau=0$ which corresponds to a 
point $(h,\theta$) lying beneath the solid separatrix will trace out a physically 
acceptable, fully attached solution for some $Q<Q^*$. The requirement that $Q\leq Q^*$ coincides with the criterion given by Moffatt \cite{moffatt1977behaviour}.
%
%%%%%%%%%%
%%%%%%%%%%%
%

In Figure~\ref{fig:omega_star}(a) we show the variation with the dimensionless flux $Q$ of the dimensionless fluid volume,
\bea
V = \int_0^{2\pi} h(\theta)\, d\theta.
\eea
The plot assumes a smooth solution with no shocks. The maximum volume, attained when $Q=Q^*$, is given by
\bea\label{maxsepVol}
V^* = 12\Omega^{1/2} \int_{\hat h_m}^1 \frac{dx}{(x+2)^{1/2}(x^3+3x-2)^{1/2}},
\eea
where %$\hat h = h/\Omega^{1/2}$ and 
$\hat h_m$ is the minimum value of $h/\Omega^{1/2}$, which 
occurs at $\theta=\pm\pi$. In the case of $\Omega = 1$, $\hat h_{\mathrm{min}}=p^{1/3} - 1/
p^{1/3}\approx 0.5961$, where $p=1+\sqrt{2}$. The integral \eqref{maxsepVol} can be 
calculated exactly but the resulting expression in terms of elliptic integrals is unwieldy and is 
not included here. 
Numerically we calculate $V^*/(2\pi\Omega^{1/2}) \approx0.70708140$, which agrees with the 
value quoted by O'Brien \& Gath \cite{o1998location}.

If at $t=0$ the starting profile $\textup{h}_0(\theta)$ in the initial condition \eqref{ic} coincides 
with one of the phase plane trajectories with $Q\leq Q^*$ in figure~\ref{fig:pplane}, $H(\theta)
$, say, then $h(\theta,t)=H(\theta)$ for all $t\geq 0$. Consider instead a starting profile $
\textup{h}_0$ that does not coincide with a steady solution such as that shown with a dashed 
red line in figure~\ref{fig:pplane}. If all or part of the profile crosses the solid gold separatrix then $h(\theta,t)$ 
will blow up at a finite-time singularity. Assume that $\textup{h}_0(\theta)$ is everywhere 
underneath the solid gold separatrix. Then $h(\theta,t)$ will be confined between the two 
osculating trajectories which are just tangent to the maximum and minimum of $\textup{h}_0$ 
(these trajectories are shown with heavier solid blue lines in figure~\ref{fig:pplane}). This is 
clear since each point on the initial profile $\textup{h}_0$ must traverse one of the level 
curves sandwiched between the two osculating curves.
During the subsequent motion, the film 
slope $g\equiv h_\theta$ satisfies the equation obtained by differentiating \eqref{moffatt}
\bea
g_t + (\Omega-h^2\cos\theta)g_\theta - 2hg^2\cos\theta + 2h^2g\sin\theta + \frac{1}{3}h^3\cos\theta = 0.
\eea
By the argument given above, $h$ is bounded, and the $g^2$ term is expected to drive the 
solution toward a finite-time slope singularity in a region where $\cos\theta>0$ (a similar 
assertion was made by Moffatt \cite{moffatt1977behaviour} via a slightly different argument).
This infinite slope singularity heralds the onset of film overturning and shock formation, a phenomenon captured by the analysis of Hinch, Kelmanson \& Metcalfe \cite{hinch2004shock}.
%, who also provided 
%the shock formation time scale $\Omega_*^2 \mu^3 a^3/\rho^3 g^3 \bar h^6$. Beyond the point of shock formation the thin-film model breaks down as it assumes that $h$ is a single-valued function of $\theta$. 
%\cbl \[
%G=\frac{\rho g a}{\mu\Omega_*}, \qquad \eps^{-6}G^{-3} = \frac{\mu^3\Omega_*^3}{\rho^3 g^3 a^3}\frac{a^6}{\bar h^6} = \left(\frac{\Omega_*^2\mu^3a^3}{\rho^3 g^3 \bar h^6}\right)\Omega_*
%\]
%So
%\[
%\frac{\eps^{-2}}{\Omega_*} = \eps^{4}G^{3}\left(\frac{\Omega_*^2\mu^3a^3}{\rho^3 g^3 \bar h^6}\right ),
%\qquad \frac{\eps^{-6}G^{-3}}{\Omega_*} = \frac{\Omega_*^2\mu^3a^3}{\rho^3 g^3 \bar h^6}.
%\]
%\cb

For the time-dependent problem it is worth recording that, with $\Omega$ constant, equation 
\eqref{maineq} possesses an infinite set of conserved quantities. We fix $\Omega = 1$ and 
define, for integer $n\geq 0$, $\chi_n = \int Q^n dh$, where $Q$ is defined in 
\eqref{qdefinition} and where we treat
$h$ and $\theta$ as being independent in the integration. Then
\bea\label{cons_quants}
\chi_n^* = \int_0^{2\pi} \chi_n \, d\th
\eea
is a conserved quantity for \eqref{maineq}. To see this, differentiate $\chi_n^*$ with respect to 
$t$ and use the fact that $h_t=-Q_{\th}$ and the $2\pi$-periodicity of $Q$ in $\th$. The case 
$n=0$ corresponds to volume conservation, but $n\geq 1$ do not have an obvious 
physical interpretation. 
We also note that with $\Omega$ constant \eqref{maineq} can be put into the Hamiltonian form
\bea \label{hamilcanon}
h_t + \frac{\partial}{\partial \theta}\left(\frac{\delta \chi_1^*}{\delta h}\right) = 0,
\eea
where $-\chi_1^*$ plays the role of the Hamiltonian.

\subsection{Stability of the steady solution}
\label{sec:steadystab}

The linear stability of the steady solutions described above for constant $\Omega$ has been 
discussed by O'Brien \cite{o2002linear} for rimming flow, and the same analysis carries over 
here. We review briefly the essential details as these will prove useful in the ensuing analysis.
 
Writing $h = h_s(\th) + \eta(\th,t)$, where $h_s(\th)$ is a steady solution of the 
MP problem for $Q<Q^*$, and $\eta(\th,t)$ is a small perturbation, we substitute into \eqref{moffatt}. Neglecting higher order terms,
\bea\label{perturbed_eq}
\eta_t + (M_s\eta)_\theta = 0,
\eea
where $M_s(\theta) = \Omega - h_s^2\cos \theta$ and, we emphasise, $\Omega$ is constant. Assuming that $Q< Q^*$ it is straightforward to show that $M>0$ (see Appendix~A).
%; on the separatrix, where $Q=2/3$, we have $M=0$ at $\theta=0$, $2\pi$).

Since the coefficients in \eqref{perturbed_eq} are $2\pi$-periodic in $\theta$, we can use Floquet theory to justify writing $\eta = \textup{e}^{i\omega t}f(\theta) + \mbox{c.c.}$, where 
$\mbox{c.c.}$ means complex conjugate, and where the $2\pi$-periodic in $\theta$ function 
$f(\theta)$ and the constant $\omega$ are to be found. Substituting into \eqref{perturbed_eq}, and 
integrating, we find $f(\theta) = c \psi(\theta)$ for arbitrary constant $c$, where
\bea \label{fsol}
\psi(\theta) = \frac{1}{M_s}\,\mathrm{e}^{-\mathrm{i}\omega\nu(\theta)}, \qquad \nu(\theta) \equiv \int_{\pi}^{\th} \frac{d\xi}{M_s(\xi)}.
\eea
The integrand in \eqref{fsol} can be expressed as the Fourier series
\bea \label{MFS}
\frac{1}{M_s(\xi)} = \sum_{n = -\infty}^{\infty}a_n\mathrm{e}^{\textup{i}n\xi},
\eea
with $a_n=\overline{a}_{-n}$. The zeroth mode has
\bea \label{a0sol}
a_0 = \frac{1}{2\pi}P(Q), % \frac{1}{2\pi} \int_0^{2\pi}\frac{d\th}{\Omega - h_s^2 \cos\th}.
\eea
where $P(Q)$ was defined in \eqref{period}.
It is clear from \eqref{fsol} that the required periodicity of $f$ is assured only if $\omega a_0=m$ for $m\in \mathbb{N}\cup \{0\}$.
Since $a_0$ is real it follows that $\omega$ is real and the MP solution 
is neutrally stable (O'Brien \cite{o2002linear}). 
%For future reference it is convenient to define the fundamental frequency
%\bea \label{ombase}
%\omega^* = 1/a_0.
%\eea
%Then all other perturbation modes have frequency an integer multiple of $\omega^*$.
This neutral stability was also previously noted by Villegas-D\'iaz {\it et al.} \cite{villegas2003stability}, who used the method of characteristics to obtain the solution of \eqref{perturbed_eq},  
\bea
\eta(\theta,t)   = \frac{1}{\Omega -  h_s^2\sin \th }\,U\!\left( t - \int_{\th_0}^\th \frac{d\xi}{M_s(\xi)}\right),
\eea
where $\theta_0$ is an arbitrary constant.
The function $U$ is set by the disturbance to the steady surface profile at $t=0$. Villegas-D\'iaz {\it et 
al.} \cite{villegas2003stability} also discussed the stability of the steady MP solution for the special 
case $Q=Q^*$.

The fact that linear perturbations are stable, taken together with the 
observation from the previous section that any perturbation from a steady 
solution will lead to a finite-time slope singularity, makes clear that nonlinearity plays an important role in the dynamics, even for arbitrarily small perturbations. In passing it is interesting to note that 
an extended form of \eqref{moffatt} that incorporates higher order terms in the lubrication 
approximation was derived by Benilov {\it et al.} \cite{benilov2003new} and was also 
shown to have neutrally stable steady solutions. Despite the 
neutrality of its eigenmodes, Benilov {\it et al.} \cite{benilov2003new} showed that the 
linearisation of the extended equation about a steady state 
admits a so-called `explosive instability': despite the linearised problem yielding an infinite number of 
bounded harmonic modes (which would normally be taken to imply stability) it supports explosive 
disturbances that blow up in finite time.

%\crd AJBL: Should be pointed out that this stability analysis is not valid when $Q=2/3$ (the value at the separatrix), because of the behaviour of the function $M(\th)$ at the origin. In this case, as noticed by Villegas-D\'iaz et al. \cite{villegas2003stability}, it is necessary to expand $M(\th)$ around $\th\rightarrow0^+$ and $\th\rightarrow2\pi^-$ and we can solve the characteristic equations in a close form...\cb \cbl MGB: clarification now included.\cb \textcolor{cyan}{AJBL: Nice, thank you!}

%%%%%%%%%%%%%%%%%%%%%%

\section{The characteristic dynamical system}\label{sec:dynamical_system}

In general a numerical approach is required to handle the case of a time-dependent angular velocity, 
$\Omega=\Omega(t)$. In this section we carry out a numerical investigation of the dynamical system 
\eqref{chareqs} with $\Omega(t)$ given by \eqref{omdefinition}. We leave any interpretation of the results in the context of the cylinder flow problem until later sections. 

%Intuitively, we might expect that if $(\th_0,h_0)$ lies below the heteroclinic orbit in figure~\ref{fig:pplane}, and if $b$ is sufficiently small then the trajectory $(\theta(t),h(t))$ will broadly follow one of the steady-solution $Q<Q^*$ orbits. If $b$ is large enough, we might expect that the trajectory may stray above the heteroclinic orbit and, consequently, lead to finite time blow-up.
The problem may be put into the form of the time-periodically perturbed Hamiltonian system 
\bea\label{hamilform}
\bm{J} \frac{d\bm{u}}{dt} = \nabla H_p + \bm{p}, \qquad 
\bm{p} = 
\begin{pmatrix}
-b\cos\sigma t \\[0.1in]
0
\end{pmatrix},
\qquad 
\bm{J}
=
\begin{pmatrix}
0 & -1\\[0.1in]
1 & 0
\end{pmatrix},
\eea
%\bea\label{hamilform}
%\frac{d}{d t}
%\begin{pmatrix}
%h \\[0.1in]
%\theta
%\end{pmatrix}
%= 
%\begin{pmatrix}
%-\frac{1}{3}h^3\sin \theta \\[0.1in]
%\Omega(t) - h^2\cos \theta
%\end{pmatrix}
%=
%\bm{J}\nabla H_p,
%\qquad
%\bm{J}
%=
%\begin{pmatrix}
%0 & 1\\[0.1in]
%-1 & 0
%\end{pmatrix},
%\eea
where $\bm{u}=(h,\theta)^T$ and $\nabla = (\partial_h, \, \partial_\theta)^T$. The Hamiltonian, $H_p$, is such that
\bea \label{hamil}
-H_p = h - \frac{1}{3} h^3\cos \th.
\eea
The form of $\Omega(t)$ is given in \eqref{omdefinition}.
When $b=0$ the perturbation vanishes, $\bm{p}=\bm{0}$, and the system \eqref{hamilform} is integrable. This corresponds to the steady flow MP problem discussed in section \ref{sec:steady}.

We integrate \eqref{hamilform} numerically using the Matlab routine \verb!ode45!. The initial condition is taken to be
\bea \label{blowup_ic}
\theta = -\frac{\pi}{2}, \qquad h = \textup{h}^*
\eea
at $t=\tau=0$. We build up a behavioural map of the solution space distinguishing between 
solutions that remain bounded and those that blow up in finite time.
The latter is detected by testing when $h$ exceeds a selected value.
This is sufficient to give an accurate picture: blow-up is initiated when the solution 
trajectory of \eqref{hamilform} latches onto the unstable manifold (leftmost dashed part of the 
gold separatrix in figure~\ref{fig:pplane}) and it occurs very rapidly thereafter and on a much 
shorter timescale than the time period of the cylinder oscillations. In our computations we deemed 
blow-up to have occurred when $h\geq 5.0$. 

\begin{figure}[!t]
\centering\includegraphics[width=5.2in]{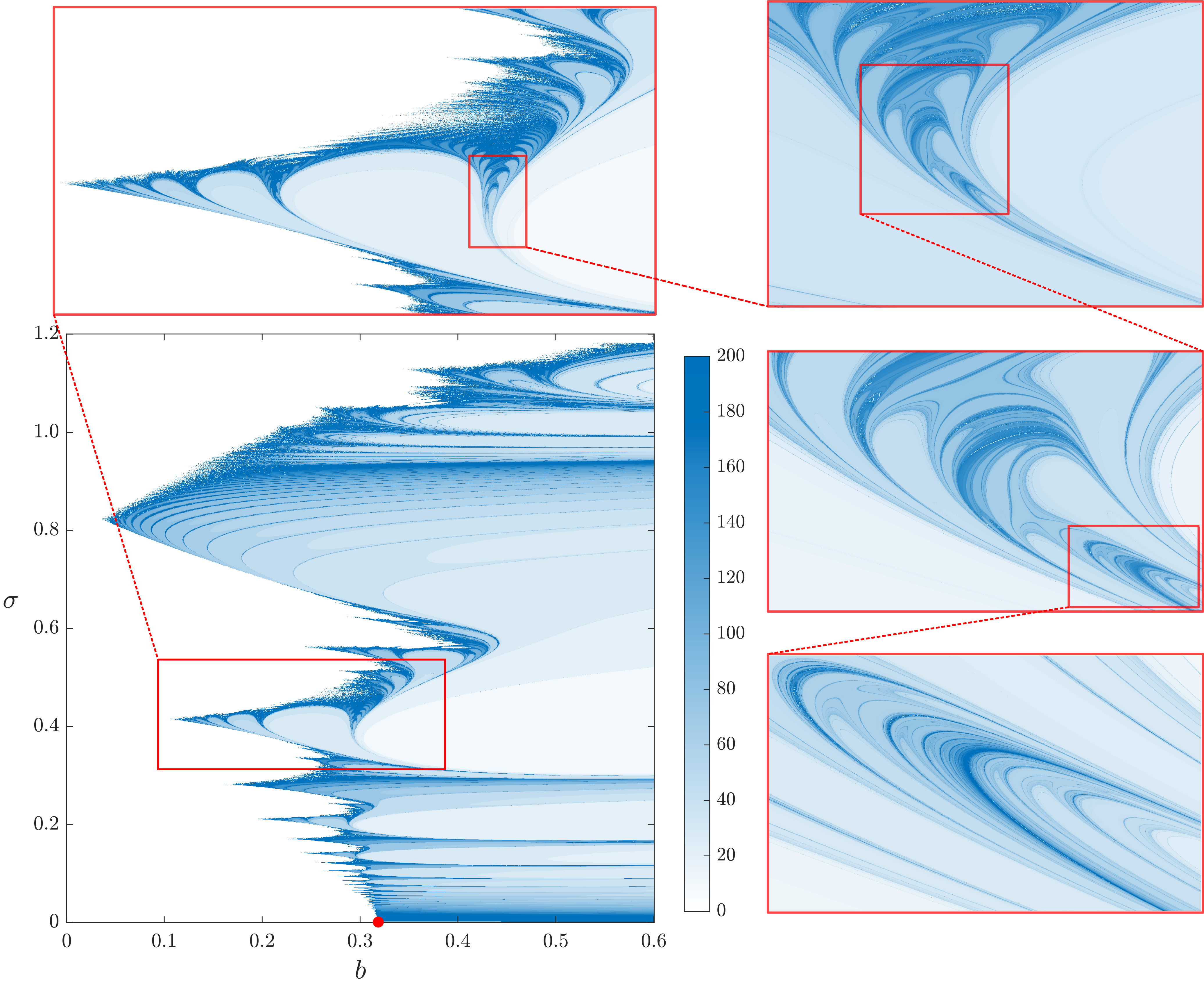}
\caption{The blow-up map obtained by integrating the characteristic equations \eqref{chareqs} 
with the initial condition \eqref{blowup_ic} taking $\textup{h}^*=1/\sqrt{3}$, with $
\Omega(t)$ given by \eqref{omdefinition}.
Points in the $b\sigma$-plane are coloured white if the solution remains bounded occurs and 
blue if blow-up occurs. The shade of blue is determined by the time at which blow-up occurs 
indicated in the colour bar: darker blue corresponds to later blow-up times. Here blow-up is 
deemed to have occurred when $h=5.0$ is reached. The red marker point at $\sigma,b)=(0,0.317)$ indicates the predicted blow-up threshold from the small-$\sigma$ analysis of section~\ref{sec:lowfreq}.}
%If we take $Q=1/\sqrt{3}$ we compute $\omega^*=0.8767$.\cb}
\label{fig:blowup_map_zooms}
\end{figure}

 Figure~\ref{fig:blowup_map_zooms} shows the behavioural map in the $b\sigma$-plane for 
 $\textup{h}^*=1/\sqrt{3}$.
The blue shading indicates the blow-up time with 
darker blue corresponding to later blow-up times. In the white region no blow-up was encountered 
in $0\leq t\leq 3500$, and we take this to mean that the solution remains regular and bounded.
%The number of the periods depends on the value of $\sigma$ from $\sim 600$ for $\sigma = 1.2$ to $\sim 5$ for $\sigma = 0.01$.}
%\\[0.1in]
Since the chosen $\textup{h}^*$ is below $Q^*=2/3$, the solution at $b=0$ traverses 
one of the level curves below the solid gold separatrix in figure~\ref{fig:pplane} and is 
therefore bounded.  The complexity of the map is 
apparent and some of the features are reminiscent of the complex structures created by 
discrete one-dimensional maps, including highly intricate boundaries and the emergence of 
apparent self-similarity upon zooming in to certain parts of the picture. The latter phenomenon 
is seen in the 
various subsidiary panels in figure~\ref{fig:blowup_map_zooms}.  

\begin{figure}[!t]
\hspace{-1in}(a) \hspace{2.5in} (b) 
\\[0.05in]
\centering\includegraphics[width=5.in]{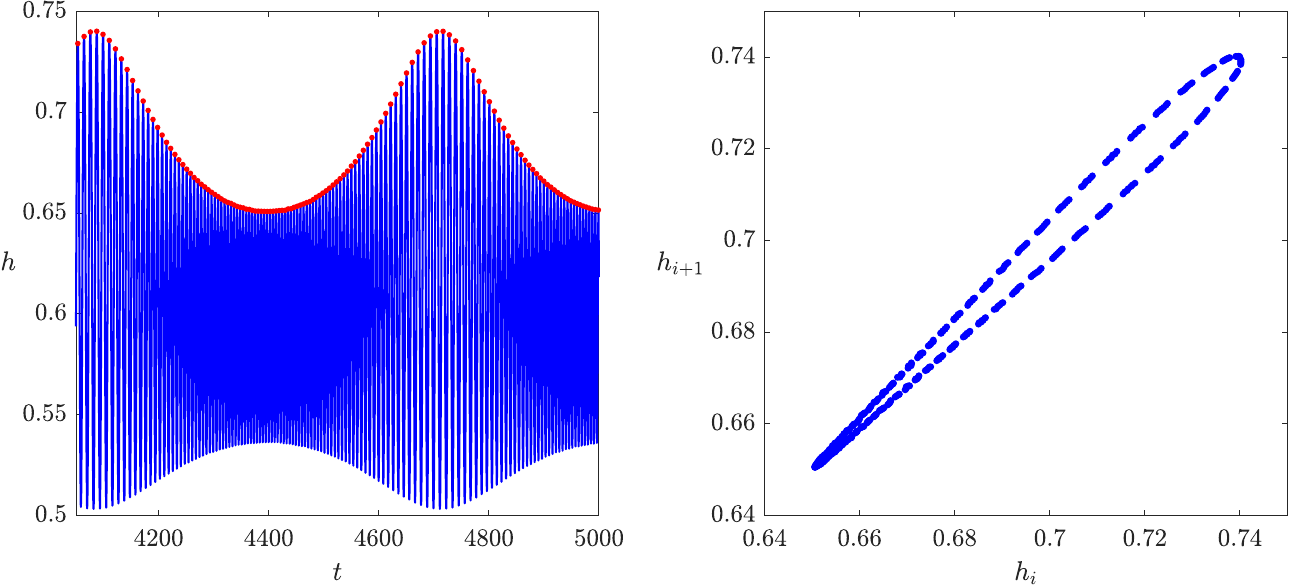}
\caption{Integration at the point $(b,\sigma)=(0.25,0.01)$ in figure~\ref{fig:blowup_map_zooms} up to $t=5000.0$. (a) Time signal over the later stages of the integration; the red markers indicate the local maxima. (b) The return map $(h^m_i,h^m_{i+1})$, where $h^m_i$ is the $i$th local maximum of the time signal.}
\label{fig:quasi}
\end{figure}

The red marker point at $(\sigma,b)=(0,0.317)$ indicates the threshold for blow-up predicted by the small-$
\sigma$ analysis to be discussed in section~\ref{sec:lowfreq}. When $\sigma$ is small the solution is found to 
be quasiperiodic in nature. This is illustrated in figure~\ref{fig:quasi} for the point $(b,\sigma)=(0.25,0.01)$ in 
figure~\ref{fig:blowup_map_zooms}. The integration was carried out up to $t=5000.0$. Panel (a) shows the 
time signal over the last $1000$ time units, and panel (b) shows a return map with $(h^m_i,h^m_{i+1})$, 
where $h^m_i$ is the $i$th local maximum of the time signal, shown with dots. The appearance of an almost complete closed 
loop in the return map is the classic hallmark of 
quasiperiodic dynamics (e.g. Guckenheimer \& Holmes 
\cite{guckenheimer_holmes_2002}). 

A particularly interesting 
feature of the map is that there appears to be some sort of resonance manifesting as a sequence of 
sharp protrusions extending to the left. With $\textup{h}^*=1/\sqrt{3}$ the unforced oscillator (viz 
\eqref{hamilform} with $\bm{p}=\bm{0}$) has the natural
frequency $\omega^*(1/\sqrt{3}) =  0.877$. The uppermost protrusion in 
figure~\ref{fig:blowup_map_zooms} has its apex at a value of the forcing frequency, $\sigma$, that is 
close to this. Moreover, the various protrusions below occur at values of $\sigma$ that are close to 
rational multiples of $\omega^*$. The fact that these apparently resonant values of $\sigma$ do not 
quite coincide with $\omega^*$ (or a rational multiple thereof) is presumably attributable to the 
nonlinearity of the underlying oscillator. 

%
%%%%%%%%%%
%%%%%%%%%%%
%
%%%%%%%%%%
%
% Figure data in 
%
% Matlab/characteristics/Origin_new/z-results/returnmaps/sig=0p01_b=0p25/
%
%%%%%%%%%%%
%

\cb

\section{The Moffatt-Pukhnachev flow with periodic modulation}\label{sec:MPF_modulation}

Since the original partial differential equation, \eqref{maineq}, for the cylinder flow problem is 
hyperbolic, the appearance of shocks is expected. Therefore, while indicative, the 
results of the previous section, which followed single-trajectory solutions of the Hamiltonian 
system \eqref{hamilform}, should be interpreted with some care in the context of the rotating cylinder problem.
In this section we study solutions to the initial value problem \eqref{maineq}, \eqref{ic} with $\Omega(t)$ given in \eqref{omdefinition}.
It will be important to distinguish between initial conditions that correspond to 
solutions of the equivalent instantaneous steady problem and those that do not. To this end 
we define the class of functions
\bea
\mathscr{H} = \{h(\theta): \,\, \Omega_0 h - \frac{1}{3} h^3\cos\th = \mathcal{Q}_0 \,\,\mbox{for some} \,\, \mathcal{Q}_0\in[0,2/3] \},
\eea
where $\Omega_0=\Omega(0) = 1+b$. If we write $h=\Omega_0^{1/2}H$ and 
$\mathcal{Q}_0=q_0\Omega_0^{3/2}$ then the restriction in $\mathscr{H}$ reduces to 
$H - \frac{1}{3} H^3\cos\th = q_0$, contours of which correspond to solutions of the steady 
MP problem and are shown in figure~\ref{fig:pplane}.
%
%%%%%%%%%%
\begin{figure}[!t]
\hspace{-2in} (a) 
\\[0.1in]
\centering \includegraphics[width=4.25in]{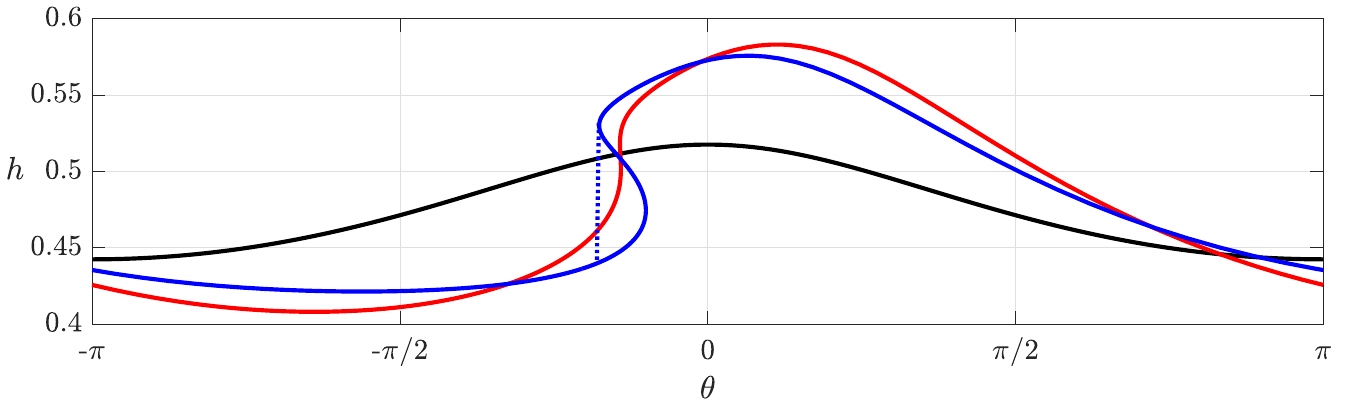}
\\
\hspace{-4.25in} (b) 
\\[0.1in]
\centering \includegraphics[width=4.25in]{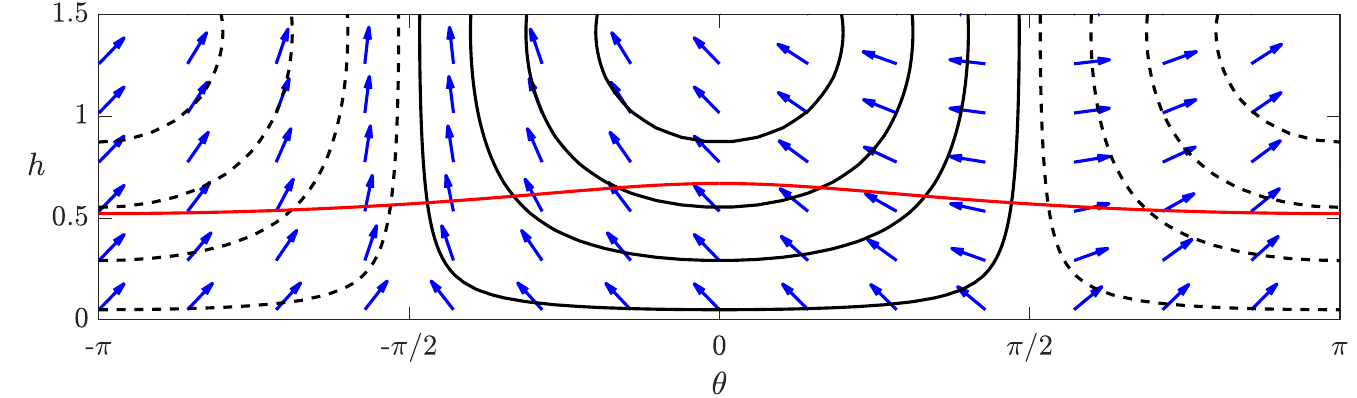}
\caption{(a) Typical solution behaviour for the model equation \eqref{maineq} in the presence 
of modulation, here with $b=0.25$ and $\sigma=0.5$. The computation was carried out by 
solving the characteristic system \eqref{hamilform} numerically using Matlab routine ode45. 
The initial profile (black curve) $\textup{h}_0 \in \mathscr{H}$ has 
$\mathcal{Q}_0 = \sqrt{2}/3$. The 
profile at the occurrence of the slope singularity at $t \approx 55.4$ is shown in red. The blue 
curve shows the film profile predicted by 
\eqref{hamilform} at $t = 60$; the blue dotted line indicates how a shock could be introduced 
to interpret the profile as a single-valued solution of \eqref{maineq}.
(b) The strain and vorticity fields for the flow with streamfunction $\mathcal{H}_p$ given in 
\eqref{hamil}. The arrows show the dominant eigenvector of the associated rate of strain 
tensor and the continuous lines show contours of constant positive (solid lines) and negative 
(dashed lines) vorticity. The solid red line shows a steady solution (see figure~\ref{fig:pplane}) 
solving  \eqref{moff2} with $\Omega=1$ and $Q=0.57$.
}
\label{fig:overturning}
\end{figure}
%%%%%%%%%%%
%

Intuitively we might expect the $b\neq 0$ oscillations to cause the film 
surface to overturn and become multi-valued signifying the breakdown of \eqref{maineq}. This 
is indeed the case, and for a general choice of modulation frequency $\sigma$ and amplitude $b$ the solution $h(\theta,t)$ develops an infinite slope singularity at some time $t^*>0$. Sample profiles are shown in figure~\ref{fig:overturning} for the 
case $\mathcal{Q}_0 = \sqrt{2}/3$ 
(black curve) taking 
$b=0.25$ and $\sigma=0.5$. The numerical computations were carried out by integrating the 
characteristic system \eqref{hamilform} forward in time from the starting profile using the Matlab routine \verb!ode45!. A parametric representation is introduced in which $(\theta(\xi_i,t),h(\xi_i,t))$ is tracked in time for $i=1,\ldots,N$, where $\xi_i$ is one of $N$ equally-spaced points in the range $[0,2\pi)$.
The onset of overturning is detected by monitoring if there is a sign change in $
\theta_\xi$, the differentiation in $\xi$ being done with spectral accuracy using a FFT. Typically we 
found that using a grid with $128$ equally-spaced points in $\xi$ is sufficient to get an accurate 
solution; the Matlab integrator \verb!ode45! uses an adaptive time step that, in the results to be 
presented, typically varies between $10^{-3}$ and $10^{-5}$.

%
%%%%%%%%%%
\begin{figure}[!t]
%
% Figure {b0p3_q0p5.eps} created by MGB using matplotlog.m in:
% Matlab/characteristics/Origin_new/z-results/CaseA-b=0p3_q0=0p5_discontinuity/
% New_MGB_runs/Omega_not_1
%
% Figure {b0p3_q0p5_Om1.eps} created by MGB using matplotlog.m in:
% Matlab/characteristics/Origin_new/z-results/CaseA-b=0p3_q0=0p5_discontinuity/
% New_MGB_runs/Omega=1
%
% Figure {b0p5_q_sqrt2o3.eps} created by MGB using matplotlog.m in:
% Matlab/characteristics/Origin_new/z-results/CaseB-b=0p5_q0=0p4714/
%
%
\centering
\hspace{-2.35in} (a) \hspace{2.65in} (b) \hspace{4in}
\\[0.05in]
\includegraphics[width=2.65in]{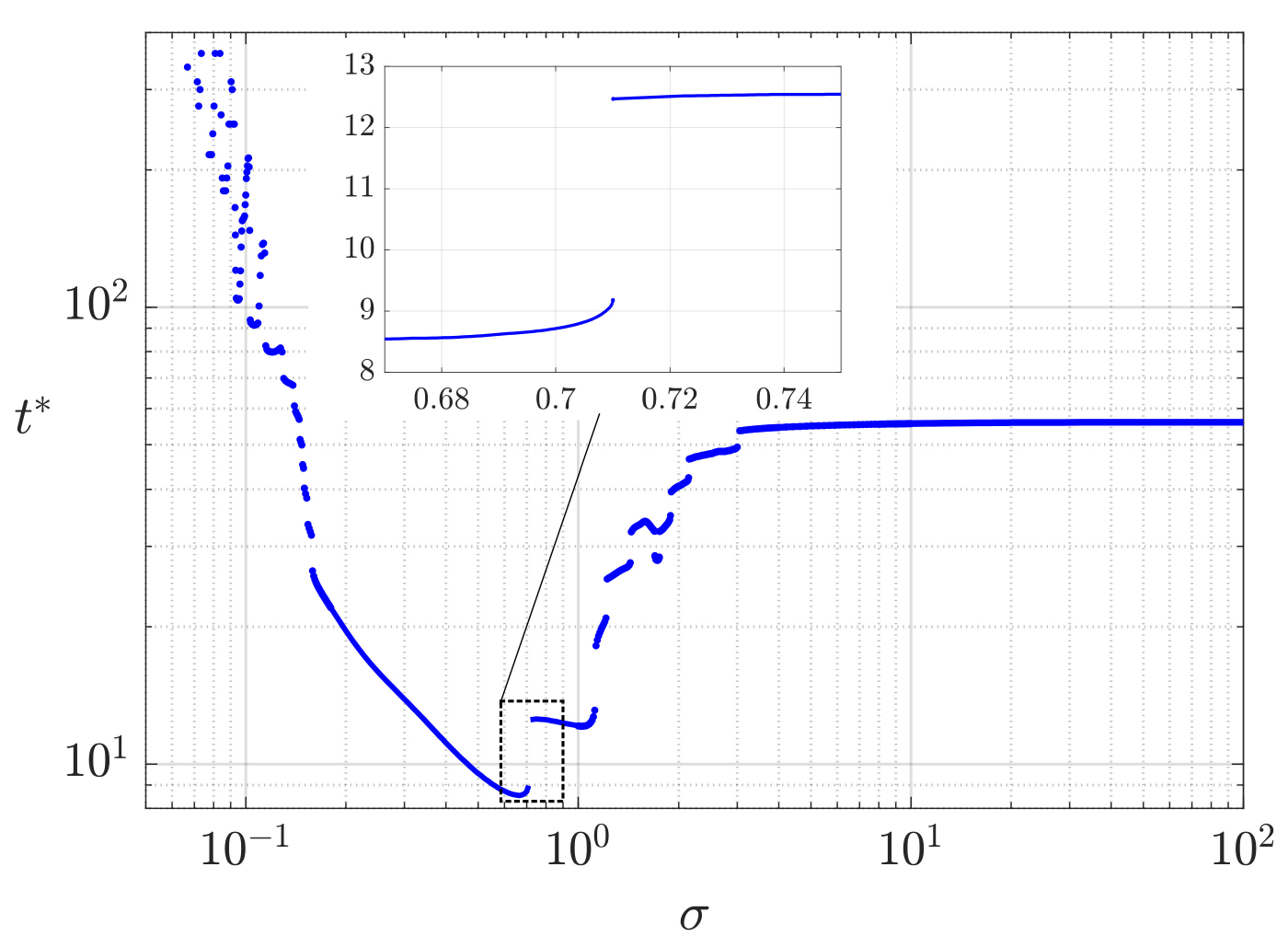}
\includegraphics[width=2.6in]{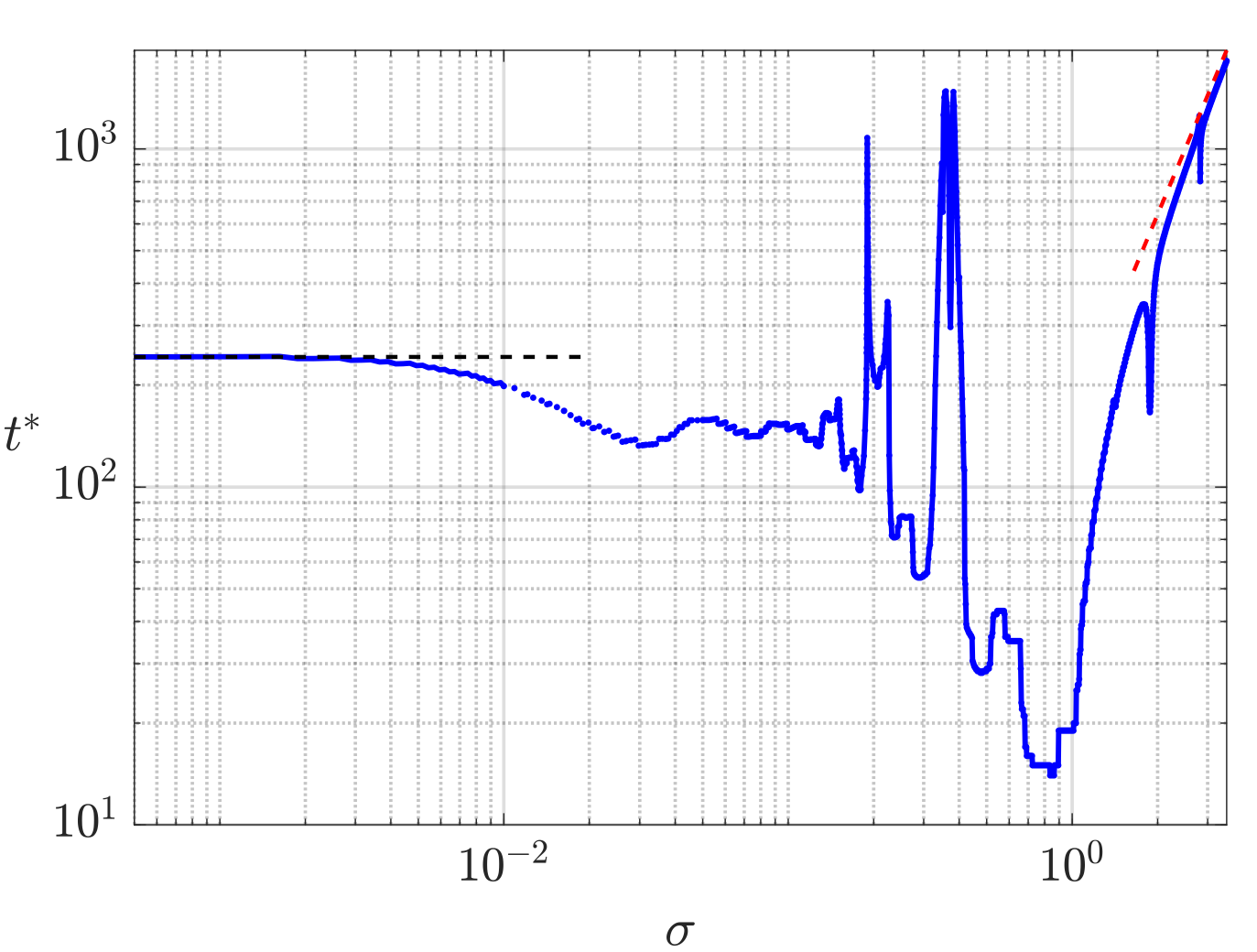}
\\[0.1in]
%\hspace{-2.5in} (c) 
\hspace{-2.35in} (c) \hspace{2.65in} (d) \hspace{4in}
\\[0.05in]
\includegraphics[width=2.6in]{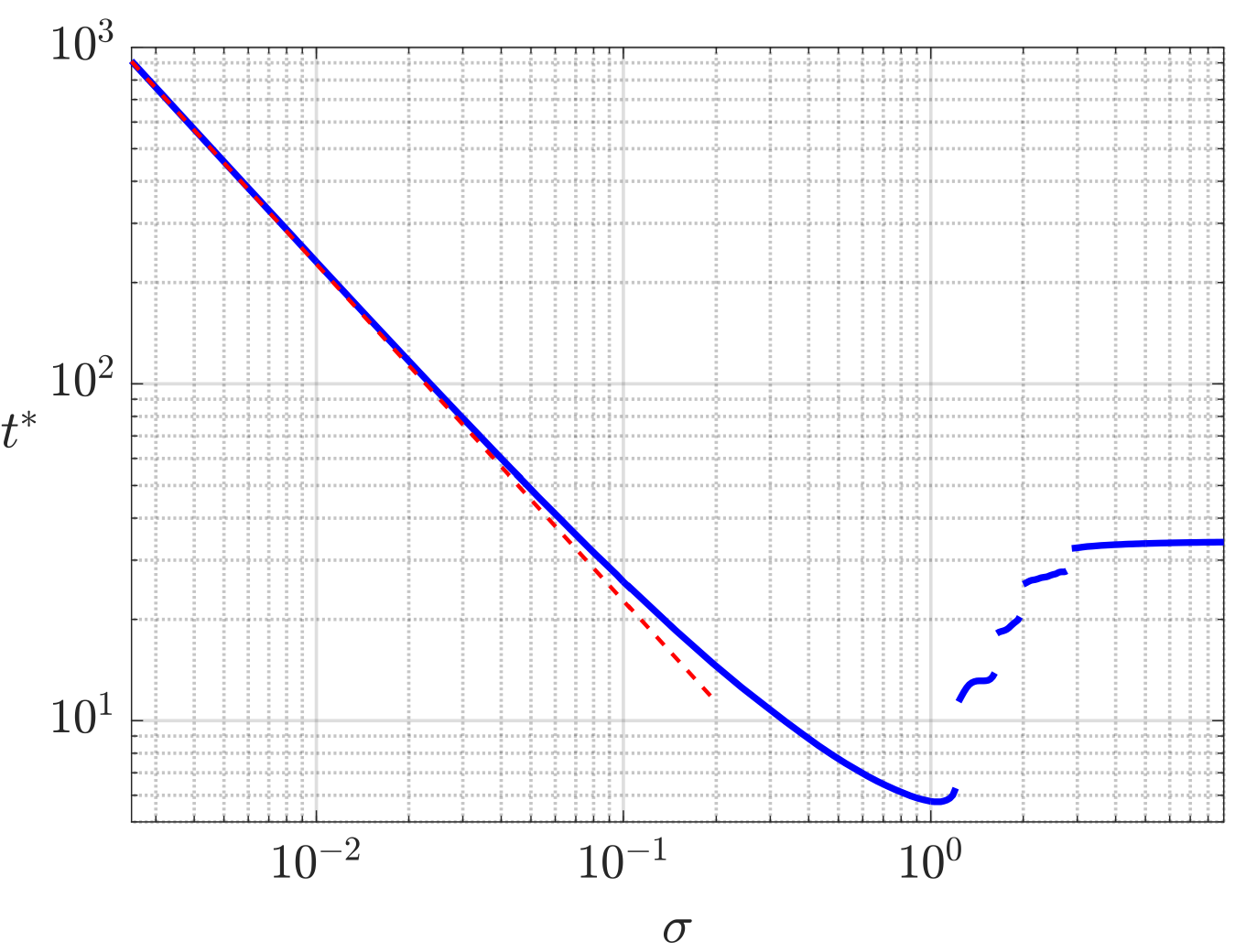}
\includegraphics[width=2.6in]{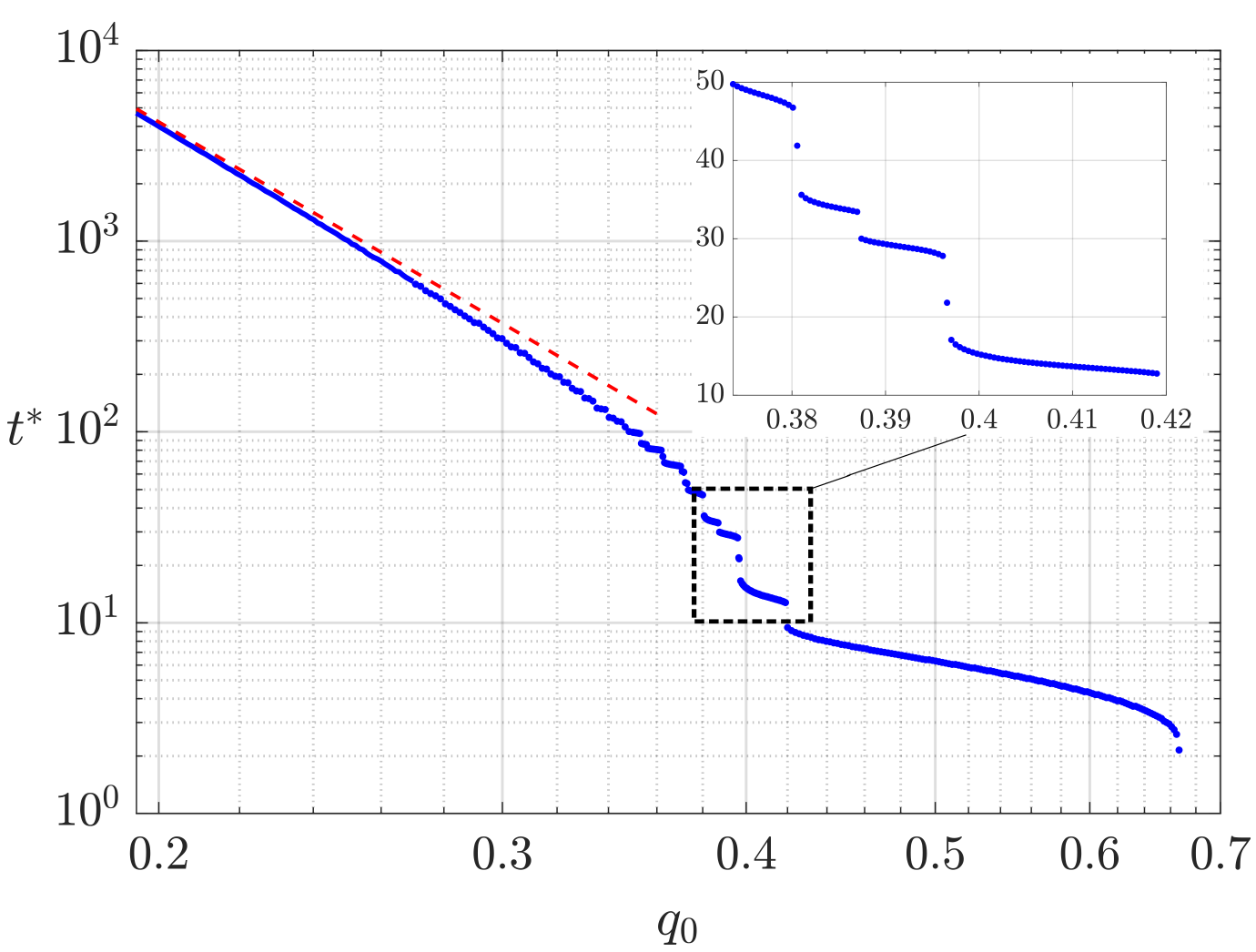}
\caption{The overturning time, $t^*$, shown against modulation frequency $\sigma$, for
the governing equation \eqref{maineq} with initial condition \eqref{ic} and starting profile
$\textup{h}_0$.  (a) $b=0.3$, $\textup{h}_0(\theta)\in \mathscr{H}$ with
$\mathcal{Q}_0=q_0\Omega_0^{3/2}$, $q_0=0.5$, (b) $b=0.3$, $\textup{h}_0\notin \mathscr{H}$ solving
$\textup{h}_0 - (1/3)\textup{h}_0^3\cos\theta = 1/2$, and (c) $b=0.5$, $\textup{h}_0\in \mathscr{H}$
with $\mathcal{Q}_0=q_0\Omega_0^{3/2}$, $q_0=\sqrt{2}/3=0.471$.
For (a) the point $(b,q_0)=(0.3,0.5)$ lies in the white region in 
figure~\ref{fig:secularity_cond} corresponding to a regular asymptotic solution in the small-$\sigma$ limit (refer to section~\ref{sec:lowfreq}). In (b) the black dashed line indicates the 
overturning time obtained by integrating \eqref{smallsigexp} with $\Omega(T)$ frozen at $\Omega_0$; the red dashed 
line shows the curve $t^*=160\sigma^2$, the coefficient having been chosen to obtain a 
reasonable fit. In (c) the point $(b,q_0)=(0.5,0.471)$ lies in the blue blow-up region in 
figure~\ref{fig:secularity_cond}, and the red dashed line shows the curve $t^*= T_s^*\sigma^{-1}$, where $T_s^* = 2.270$ is computed by integrating \eqref{qeq} subject to 
$q(0)=\sqrt{2}/3$. (d) The overturning time $t^*$ versus $q_0$ for $b = 0.5$ and $\sig = 0.6$. 
The initial condition is \eqref{ic} with $\textup{h}_0\in \mathscr{H}$ and $\mathcal{Q}_0= q_0 
\Omega_0^{3/2}$, where $\Omega_0=\Omega(0)$; the red dashed 
line shows the curve $t^*=0.27q_0^{-6}$, the coefficient having been chosen to obtain a 
reasonable fit.
%\crd {\bf We should note somewhere that $t^*$ depends also on the phase of $\Omega$.}\cb
}
\label{fig:Tstar_mgb}
\end{figure}
%%%%%%%%%%%
%
In figure~\ref{fig:Tstar_mgb}(a-c) we show how the overturning time $t^*$ varies with $\sigma$
for two cases chosen so that $\textup{h}_0\in\mathscr{H}$ in one and $\textup{h}_0\notin\mathscr{H}$ 
in the other. The modulation amplitude is set at $b=0.3$. The asymptotic 
results shown in this figure will be discussed in a later section. It is quite striking, particularly in 
figure~\ref{fig:Tstar_mgb}(a), that the $t^*(\sigma)$ curve exhibits numerous jump 
discontinuities. The presence of these discontinuities has 
been confirmed by meticulous numerical computations, including a very careful convergence 
study using up to $N=1024$ points in $\xi$ and a time step of size $10^{-5}$. 
%\crd Here it would be good to restore Antonio's inset in figure~\ref{fig:Tstar_mgb}(a) to show the gap in 
%close-up. \cb {\bf \textcolor{cyan}{AJB: If you want, you can send to me the figure as a .fig file 
%and I could put the inset in it.} }

% This gap is located around $\sig\approx0.71$. The appearance of the gaps seems to be related with the two different mechanisms of overturning.\\

To explain the jump discontinuities, and with reference to the system \eqref{hamilform}, we 
regard the Hamiltonian $H_p(\theta,h)$ given in \eqref{hamil} as the stream 
function for a two-dimensional incompressible fluid flow in the 
$\theta h$-plane. In this interpretation the flow occupies the whole plane and is not confined to the 
domain of the liquid for the cylinder problem. Solving \eqref{hamilform} elicits the trajectories 
of individual particles advected passively within this flow, with the liquid surface corresponding 
to a material line. As it moves with the flow, this material line is stretched and distorted by the flow's
strain and vorticity fields. These fields, which are 
independent of time, are illustrated in figure~\ref{fig:overturning}(b): the arrows show the 
direction of dominant strain corresponding to the eigenvector of the rate of strain tensor with 
positive eigenvalue (we recall that since the rate of strain tensor is symmetric its two 
eigenvalues, which are real, sum to zero, and its two eigenvectors are mutually orthogonal).
In general, material elements tend to align with the dominant strain eigenvector whilst being 
rotated one way or another according to the sign of the vorticity. The constant vorticity 
contours are solid where the vorticity is positive, promoting counterclockwise rotation, and 
dashed where the vorticity is negative, promoting clockwise
rotation. (For $b=0$ a steady solution has a surface profile that stays fixed as the competitive 
effects of strain and rotation are in perfect balance; see the solid red line in 
figure~\ref{fig:overturning}b.)
 
Certain parts of the strain-vorticity field promote steepening of the material surface and others 
promote flattening. The region around $\theta=-\pi/2$ presents a particular danger zone for 
surface steepening. Here the magnitude of the vorticity is small and the dominant strain 
eigenvectors are almost vertical. In contrast, 
the region around $\theta=\pi/2$ has weak vorticity and almost horizontal strain vectors, and 
therefore this region strongly encourages flattening of the surface. It appears that we get a 
jump discontinuity in $t^*$ at a certain $\sig$ because, just below this frequency, 
the surface overturns near to $\theta=-\pi/2$; but for a slightly larger $\sigma$  the 
overturning is just avoided and the surface must travel a further distance through a less 
dangerous region before overturning is finally induced in the danger zone some time units later.

In figure~\ref{fig:Tstar_mgb}(d) we show the overturning time $t^*$ for the case $b = 0.5$ and $\sig=0.6$ and 
for a range of different initial conditions. The latter are given by \eqref{ic} with $\textup{h}_0\in \mathscr{H}$ 
and $\mathcal{Q}_0= q_0 \Omega_0^{3/2}$. The panel shows $t^*$ plotted against $q_0$, and we see the 
occurrence of jump discontinuities at certain values of $q_0$. Some of 
these are highlighted in the inset. Since $q_0\to 0$ corresponds to taking an initial profile that is close to the 
wall and almost flat (see figure~\ref{fig:pplane}), we expect the overturning time to diverge in this limit. The 
dashed red line in figure~\ref{fig:Tstar_mgb}(d) suggests that this is indeed what occurs and, moreover, it 
happens such that $t^* \sim q_0^{-6}$. 

To further understand the dynamics it is instructive to consider the flow in the high frequency ($\sigma\gg1$) and low frequency ($\sigma\ll 1$) limits. These are examined in the following subsections.

\subsection{High frequency limit ($\sigma\gg 1$)}

Our numerical results suggest that the flow is periodic in time if $\sigma$ is sufficiently large, and this motivates an investigation of the dynamics when $\sigma \gg 1$. In this limit 
there are two naturally disparate time scales in the problem: an $O(1)$ time scale
associated with the steady part of $\Omega$, that is the continuous rotation, and an $O(1/\sigma)$ 
timescale associated with the rapid superimposed oscillations. Having said this, we might intuit that for 
an initial condition corresponding to a steady Moffatt-Pukhnachev solution, the former timescale is essentially removed and the flow will develop on timescales of the rapid oscillation. 

Keeping the preceding comments in mind, we perform a multiple-scales analysis incorporating both time scales. Let $h = h(\th,t,T)$ with $T = \sigma t$ acting as the fast time variable and $t$ acting as the slow time variable. Assuming in the usual way that $t$ and $T$ are independent \eqref{maineq} becomes
%\bea \label{multiple_scales}
%h_t + \sigma h_T + \left(\Omega(T) -  h^2\cos\th\right) h_\th + \frac{1}{3} h^3\sin\th = 0,
%\eea
\bea \label{multiple_scales}
h_t + \sigma h_T + \left(\Omega(T)h -  \frac{1}{3}h^3\cos\th\right)_\th  = 0,
\eea
where $\Omega(T) = 1 + b\cos T$.
We expand by writing
\bea \label{largesigexpand}
h(\th,t,T) = h_0(\th,t,T) + \sigma^{-1} h_1(\th,t,T) + \sigma^{-2} h_2(\th,t,T) + \sigma^{-3} h_3(\th,t,T) + O(\sigma^{-4}).
\eea
Introducing this expansion into \eqref{multiple_scales}, at $O(\sigma)$ we find that $h_{0T} = 0$, which implies $h_0 = h_0(\th,t)$. At order $O(1)$ we obtain 
$h_{1T} = - bh_{0\th}\cos T - F(\theta,t)$, where
\bea\label{order1}
F(\theta,t) = h_{0t} + \left( h_0 - \frac{1}{3} h_0^3\cos\th\right)_\theta.
\eea
Integrating with respect to $T$,
\bea \label{h1sol0}
h_1(\th,t,T) = - bh_{0\th}\sin T - F(\theta,t)T + A_1(\th,t),
\eea
where $A_1(\th,t)$ is an arbitrary function of integration. In order to avoid secular terms, we demand that $F = 0$. Then the leading order term $h_0$ satisfies the original equation \eqref{maineq} with constant unit forcing frequency, $\Omega=1$, that is it corresponds to
a solution of the constant rotation rate Moffatt-Pukhnachev problem.
To prevent $h_0(\theta)$ developing a slope singularity (see section~\ref{sec:steady}), we take $h_0\equiv h_s(\theta)$, where the steady solution $h_s$ satisfies \eqref{moff2} with $\Omega=1$ for some flux $Q<2/3$.

Proceeding, we have
\bea\label{h1sol}
h_1(\th,t,T) = - bh_{0\th}\sin T  + A_1(\th,t).
\eea
At order $O(\sigma^{-1})$,
\bea\label{sigma_minus_1}
%h_{1t} + h_{2T} +\left(\Omega(T)h_1 -%h_0^2h_1\cos\th\right)_\th = 0,
h_{2T} = -h_{1t} - \left((\Omega(T) - h_0^2\cos\th ) h_1 \right)_\th.
\eea
Substituting \eqref{h1sol} into the right hand side of \eqref{sigma_minus_1}, we see that
secular terms in $h_2$ will not arise if the $T$ independent terms vanish, that is if
\bea\label{A1eq}
A_{1t}+\left(\mathcal{M} A_1\right)_\th = 0,
\eea
where $\mathcal{M} = 1-h_0^2\cos\th$.
This is the same as the equation that governs the linear stability of the MP problem, namely \eqref{perturbed_eq}.
Since $h_0$ has been chosen to be a steady solution of the MP problem, 
%\eqref{A1eq} is analogous to %\eqref{perturbed_eq} 
according to the results of section~\ref{sec:steadystab}, $A_1$ is $t$-periodic with frequency $\omega=m/a_0$, where $a_0$ is given by \eqref{a0sol} ($\Omega=1$) and $m\in \mathbb{N}\cup \{0\}$. Hence we have
\bea\label{A1sol}
A_1(\theta,t) = \lambda \ee^{i\omega t}\psi (\theta) + \mathrm{c. c.},
\eea
where $\lambda$ is an arbitrary constant. The $2\pi$-periodic function $\psi(\theta)$ was given in \eqref{fsol}.

Integrating \eqref{sigma_minus_1},
%\bea \label{h2sol}
%h_2 = -\frac{1}{4}b^2h_{0\th\th}\cos 2T - bh_{0\th\th}\cos T - bA_{1\th}\sin T + b\cos T\left(h_0^2h_{0\th}\cos\th\right)_\th + A_2(t,\th),
%\eea
%\bea \label{h2sol}
%h_2 = -\frac{1}{4}b^2h_{0\th\th}\cos 2T + b\cos T \left(  \left(h_0^2h_{0\th}\cos\th\right)_\th  - h_{0\th\th}  \right) - bA_{1\th}\sin T  + A_2(t,\th),
%\eea
%\bea \label{h2sol}
%h_2 = -\frac{1}{4}b^2h_{0\th\th}\cos 2T - b \left( \mathcal{M}_\theta h_{0\th} + \mathcal{M}h_{0\th\th} \right)\cos T - bA_{1\th}\sin T  + A_2(\th,t),
%\eea
\bea \label{h2sol}
h_2 = -\frac{1}{4}b^2h_{0\th\th}\cos 2T - b ( \mathcal{M} h_{0\th})_\th \cos T - bA_{1\th}\sin T  + A_2(\th,t),
\eea
where $A_2(\th,t)$ is an arbitrary function of integration.
At order $O(\sigma^{-2})$
%\bea\label{sigma_minus_2}
%h_{3T} = -h_{2t} + \left(\Omega(T) + h_0^2\cos\th\right)h_{2\th} + \left(2h_0h_1\cos\th\right)h_{1\th} + \\ \nonumber \left(h_1^2+2h_0h_2\right)h_{0\th}\cos\th -\left(h_0h_1^2 + h_0^2h_2\right)\sin\th.
%\eea
%\bea\label{sigma_minus_2}
%h_{3T} = -h_{2t} - \left( \Omega(T)h_2 - h_0 ( h_0h_2 + h_1^2 ) \cos \th \right)_{\th}.
%\eea
\bea\label{sigma_minus_2}
h_{3T} = -h_{2t} - \left( (\Omega(T) - h_0^2\cos\th)h_2 - h_0 h_1^2 \cos \th \right)_{\th}.
\eea
The secularity condition requiring that the $T$-independent terms on the right hand side of \eqref{sigma_minus_2} be eliminated takes the form
\bea\label{A2_eqn}
A_{2t} + \left(\mathcal{M}A_2\right)_\th = \frac{dS}{d\theta} + \lambda^2 \ee^{2i\omega t} \left(\psi^2 h_0 \cos \theta \right )_{\th} + \mathrm{c .c.},
\eea
where 
\bea
S(\theta) = \frac{1}{2}b^2\! \left( h_0( h_0 h_{0\th})_{\th} \cos\th - h_{0\th\th}  - h_0^2h_{0\th}  \sin\th \right).
\eea
We seek a solution in the form
$A_{2}(\theta,t) =  \alpha_{20}(\theta) + \ee^{2\mathrm{i}\omega t}\alpha_{22}(\theta) + \mbox{c. c.}$, with 
$\alpha_{20}$ and $\alpha_{22}$ required to be $2\pi$-periodic in $\theta$. It is clear from
\eqref{A2_eqn} that $\alpha_{20}(\th)$ has this property. 
Solving for $\alpha_{22}(\theta)$ using an integrating factor, we find that it is $2\pi$-periodic if
\bea
\lambda^2 \int_{0}^{2\pi} (\psi^2 h_0 \cos\th)_\th \ (\psi \mathcal{M})^2  d\theta = 0.
\eea
This holds if $\omega=0$ (so that $\psi \propto 1/\mathcal{M}$) in which case $A_{1}=\kappa_1/\mathcal{M}(\theta)$ and $A_2=(S(\theta) + \kappa_2)/\mathcal{M}(\theta)$ for arbitrary constants $\kappa_1$ and $\kappa_2$. Thus the slow timescale $t$-dependence drops out to the current order of approximation, as was anticipated.

%
%
%%%%%%%%%%%%%%%%%%%%%%%%%%%%%%%%%%%%%%%%%%
%
% Figure created by MGB using paperplot.m in folder:
% Research-PhD/4-Antonio/Research/Matlab/characteristics/Origin_new/z-results/
% large_sigma_computations
%
\begin{figure}[!t]
\centering\includegraphics[width=4.25in]{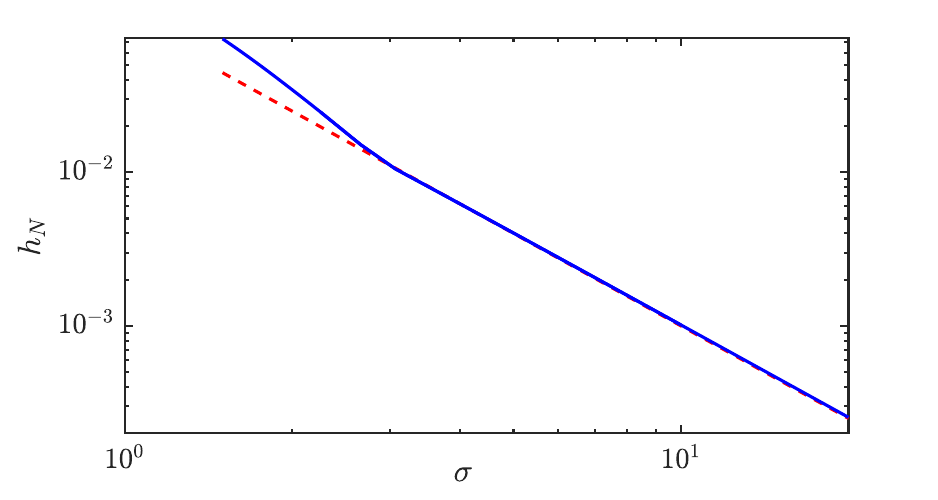}
\caption{Log-log plot of the norm $h_N$ (blue solid line) defined to be the maximum value of $h(\theta,t) - h_s(\th + \sigma^{-1} \Omega_T)$ over $\theta \in [0,2\pi)$ during the time period $t\in [0,10/\sigma]$ (here $T=\sigma t$). The function $h(\theta,t)$ was computed by solving \eqref{hamilform} numerically with the initial profile $h(\theta,0)=h_s(\theta)$, where $h_s(\theta)$ solves \eqref{moff2} with $\Omega=1$ and $Q=0.5$. The modulation amplitude $b=0.2$. A best fit curve $h_N = 0.1/\sigma^2$ is shown with a dashed red line.}
\label{fig:hdiff}
\end{figure}
%%%%%%%%%%%%%%%%%%%%%%%%%%%%%%%%%%%%%%%%%%
%
%

In the light of the preceding discussion, and recognising that the first term on the right hand side of \eqref{h1sol} represents a translation in $\theta$ by an amount $\sigma^{-1}\Omega_T$, we may now revise the expansion \eqref{largesigexpand} to 
read
\bea \label{largesigexpand2}
h(\th,t,T) = h_s(\th + \sigma^{-1} \Omega_T) + \sigma^{-1} \frac{\kappa_1}{\mathcal{M}(\theta)} + O(\sigma^{-2}). %h_2(\theta,T) + O(\sigma^{-3}).
\eea
This represents a time-periodic asymptotic solution valid when $\sigma \gg 1$.
Suppose that we compute a numerical solution to the governing equation 
\eqref{maineq} for a particular choice of parameter values and taking $\textup{h}_0(\theta)$ in the initial condition \eqref{ic} to be a steady solution $h_s(\theta)$. Since $\Omega_T(T=0)=0$, we can fit  
\eqref{largesigexpand2} to the initial profile at $t=T=0$ up to the second order in $\sigma$ 
by setting $\kappa_1=0$.
However, at second order there is no way to choose the constant $\kappa_2$ such that $h_2(T=0)=0$. We 
therefore have a discrepancy between the numerical solution and the asymptotic  
approximation of $O(\sigma^{-2})$.
In figure~\ref{fig:hdiff} we demonstrate agreement between the numerical computation and the asymptotic 
solution \eqref{largesigexpand2} with $\kappa_1=0$. Here the numerical solution was obtained by integrating 
the characteristic system \eqref{hamilform} forward in time from the starting profile $h_s(\theta)$, where 
$h_s(\theta)$ solves \eqref{moff2} with $\Omega=1$ and $Q=0.5$. The vertical axis shows the norm $h_N$ 
defined to be the maximum value of $h - h_s(\th + \sigma^{-1} \Omega_T)$ over $\theta \in [0,2\pi)$ during 
the time integration period $t\in [0,10/\sigma]$. A best fit curve $h_N=0.1/\sigma^2$ is shown with a dashed 
red line. 

The above discussion suggests that, for large $\sigma$ at least, for a sufficiently carefully chosen initial 
condition it should be possible to obtain a time-periodic solution to \eqref{maineq}  
which does not overturn. However, overturning is expected for a general initial condition and we can use our 
theory to estimate the overturning time, $t^*$. In general the starting profile $\textup{h}_0(\theta)$ in the initial 
condition \eqref{ic} will not exactly meet the requirement that $F\equiv 0$, and so we expect transient growth 
in $T$ according to \eqref{h1sol0}. 
%As noted above, matching the 
%starting profile to the asymptotic expansion \eqref{largesigexpand} evaluated at $T=0$
%produces the mismatch $h_0 = \textup{h}_0 + O(\sigma^{-2})$. 
If we identify $\textup{h}_0$ with $h_s$, where $h_s$ solves \eqref{moff2} with $\Omega=1$, it 
follows from \eqref{order1} that $F = O(\sigma^{-2})$, and the asymptotic theory predicts transient growth so 
that the uniformity of the expansion \eqref{largesigexpand} is destroyed when $t=O(\sigma^2)$. If instead 
we identify $\textup{h}_0$ with a profile that does not solve \eqref{moff2} with $\Omega=1$, then $F=O(1)$ 
and the expansion \eqref{largesigexpand} fails when $t=O(1)$. We interpret the failure of the expansion as 
the signature of overturning. This viewpoint is supported by our numerical solutions to the full 
governing equation \eqref{maineq}. For the overturning times reported in 
figure~\ref{fig:Tstar_mgb} we see that in panel (a), for which $F=O(1)$,  $t^*$ approaches a constant as 
$\sigma \to \infty$ in agreement with the preceding remarks. 
In panel (b), for which $F=O(\sigma^{-2})$, $t^*$ grows apparently like $\sigma^2$.
Evidently the onset of overturning can be considerably delayed in the high frequency limit by a judicious choice of the initial condition.
%where on the right hand side $h_0=h_0(\theta,T=0)$.

%%%%%%%%%%%%%%%%%%%%%%%%%%%%%%%%%%%%

\subsection{Low frequency limit ($\sigma\ll 1$)}\label{sec:lowfreq}

The numerical results shown in figure~\ref{fig:blowup_map_zooms} suggest that 
when $\sigma\ll 1$ blow-up occurs when $b$ exceeds a threshold value. This motivates an analysis in the low frequency limit. As in the previous subsection, we follow a multiple scales approach, in this case with $t$ as the fast time scale and $T=\sigma t$ as the slow time scale.

It is convenient at the outset to rescale the film thickness, writing $h(\theta,t,T)=\Omega^{1/2}H(\theta,t,T)$, where $\Omega(T) = 1 + b\cos T$. We assume $|b|<1$ so that $\Omega>0$ for all $T$.
We then posit the expansion 
\bea \label{smallsigexp}
H(\th,t,T) = H_0(\th,t,T) + \sigma H_1(\th,t,T) + \sigma^2 H_2(\th,t,T) + O(\sigma^3),
\eea
assuming $\sigma \ll 1$. Inserting into
\eqref{multiple_scales} we obtain at leading 
order, $O(1)$, 
%{\color{red} Antonio, please could you check the following equation?:CHECKED}
\bea \label{smallsigH0}
H_{0t} + \Omega(T) \left(H_0 -  \frac{1}{3}H_0^3\cos\th\right)_\th = 0.
\eea
%\bea\label{smallsigH0}
%H_{0\hat t} + \left(H_0 -  \frac{1}{3}H_0^3\cos\th\right)_\th = 0.
%\eea
Following the discussion in section \ref{sec:steady}, for a general initial condition we expect that $H_0$ will reach a slope singularity at a finite time $t$. However the solution will remain regular if at $t=0$ the profile  $H_0$ coincides with a $t$-independent solution of \eqref{smallsigH0}, namely one that satisfies
%{\color{red} Antonio, please could you check the following equation?:CHECKED}
\bea\label{steady_mult_scales}
H_0 - \frac{1}{3} H_0^3\cos\th = q(T),
\eea
for some $q(T)$ to  be determined later. 
We therefore insist that $H_0(\theta,T)$ solves \eqref{steady_mult_scales}. We note that this is essentially the same as the cubic equation \eqref{moff2} for the steady MP problem, with $T$ playing the role of a parameter, and with the relationship $Q=\Omega^{3/2}q$. 
%Of particular interest is the question of %whether $Q_0(T)$ can exceed the threshold %value of $(2/3)\Omega^{3/2}$ identified in
A question of interest, then, is whether $q(T)$ can reach the threshold value of $2/3$ identified in
section~\ref{sec:steady}, and so drive the leading order solution toward blow-up.

%that Note that this cubic equation for $h_0$ is very similar to that of the steady solution with $\Omega = 1$, but in this case $\Omega$ and $Q$ are functions of the independent variable $T$. The question that naturally arises from that equation is if the solution is periodic or not once $T$ comes into the play. It could happen that the at some point, the velocity profile in the $h_0-\th$ phase plane crosses the separatrix, becoming $h_0$ unbonded. On the separatrix, when $\Omega(T)$, the criterion for the existence of steady solutions is $9/4 Q^2 = \Omega^{9/4}$. Note here, that steady solution means no explicit dependecy with respect to $t$.\\

At first order, $O(\sigma)$,
\bea\label{h1_small_sigma}
H_{1t} + \Omega(M_0 H_1)_\theta = \Omega^{-1/2}R(\theta,T), \qquad R_1\equiv -(\Omega^{1/2}H_0)_T,
\eea
where $M_0(\th, T) = 1-H_0^2\cos\th$. This essentially presents a forced version of the linear stability equation \eqref{perturbed_eq}. The solution is 
\bea
H_1 = \frac{1}{M_0} U_1(z) + \frac{1}{\Omega^{3/2} M_0}\int_\pi^\theta R_1(\xi)\,d\xi,
\eea
where the function $U_1$ is arbitrary, $z = \Omega(T)  t - \nu_0(\theta)$, and
\bea
\nu_0(\theta) \equiv \int_{\pi}^\theta \frac{d\xi}{M_0(\xi)}.
\eea

We require $H_1$ to be $2\pi$-periodic in $\theta$. To check this we 
first differentiate \eqref{steady_mult_scales} with respect to $T$, and rearrange to get 
\bea\label{haggis}
M_0 H_{0T} = q_T.
\eea 
Integrating this with respect to $\theta$, 
\bea \label{snicket}
\int_{\pi}^{\th} H_{0T}\, d\theta = \nu_0(\theta) q_T.
\eea
With reference to \eqref{MFS} and \eqref{a0sol} we notice that $\nu_0(\theta)$ is composed of $a_0\theta$ plus a periodic function of $\theta$.
Rearranging \eqref{steady_mult_scales}, using the definition of $M_0$ and integrating with respect to $\theta$, we find
\bea \label{snicket2}
\int_{\pi}^{\th} H_{0}\,d\theta =
3\nu_0(\theta) q - 2\int_{\pi}^{\theta} \frac{H_{0}}{M_0} \,d\theta. 
\eea
Using \eqref{snicket} and \eqref{snicket2} we may write
\bea \label{flump}
\int_{\pi}^\theta R_1(\xi)\,d\xi 
= \Omega^{1/2} \nu_0(\theta) \left( q_T +
\frac{3}{2}\frac{\Omega_T}{\Omega} q \right)
-\frac{\Omega_T}{\Omega^{1/2}} \int_{\pi}^{\theta} \frac{H_{0}}{M_0}\,d\xi.
\eea
If $q<2/3$ then $H_0$ satisfying \eqref{steady_mult_scales} is bounded and $2\pi$-periodic in $\theta$. The right hand side of \eqref{flump}, and hence $H_1$, has the same properties if
\bea \label{qeq}
q_T = \frac{\Omega_T}{\Omega}\left(\mathcal{H}(q) - \frac{3}{2}q\right),
\eea
where
\bea\label{H0bar}
\mathcal{H}(q) = \frac{1}{P_0}\int_{0}^{2\pi} 
\frac{H_0}{1 - H_0^2\cos\theta} \, d\theta, \qquad
P_0(q) = \int_{0}^{2\pi} \frac{1}{1-H_0^2\cos \theta}\, d\theta.
\eea
The solution for $H_1$ is then given by
\bea \label{H1sol}
H_1 = \frac{1}{M_0} U_1(z) -\frac{\Omega_T}{\Omega^{1/2}} \int_{\pi}^{\theta} \frac{H_{0}}{M_0}\,d\xi.
\eea
It is clear from \eqref{H1sol} that $H_{1\theta}$ is bounded. 
Furthermore, $H_{1T}$ is bounded for all $T$ provided that $|q_T|<\infty$ by virtue 
of \eqref{haggis}.

The solution to the problem at $O(\sigma^n)$ has the form
\bea\label{small_sig_gen_sol}
H_n = \frac{1}{M_0}\left( U_n(z) + \int_\pi^\theta R_{n}(\xi)\,d\xi \right)
\eea
with $U_n$ arbitrary, and $R_{n} = -(\Omega^{1/2}H_{(n-1)})_T + ( F_n \cos \theta)_\th/3$, where the $F_n(H_0,H_1,\ldots,H_{n-1})$ are known (for example, $F_2=3H_0H_1^2$).
Let us assume that the $U_n$ are given for $n\geq 0$, for example by setting an 
appropriate initial condition at $t=T=0$. Since, if $|q_T|<\infty$, $H_n$ and $H_{n\th}$ are both bounded for $n=0$, $1$, it is clear that the same property holds for $n\geq 2$. Given these remarks we expect the expansion \eqref{smallsigexp} to remain uniform as time increases.

\begin{figure}[!t]
\centering\includegraphics[width=5.5in]{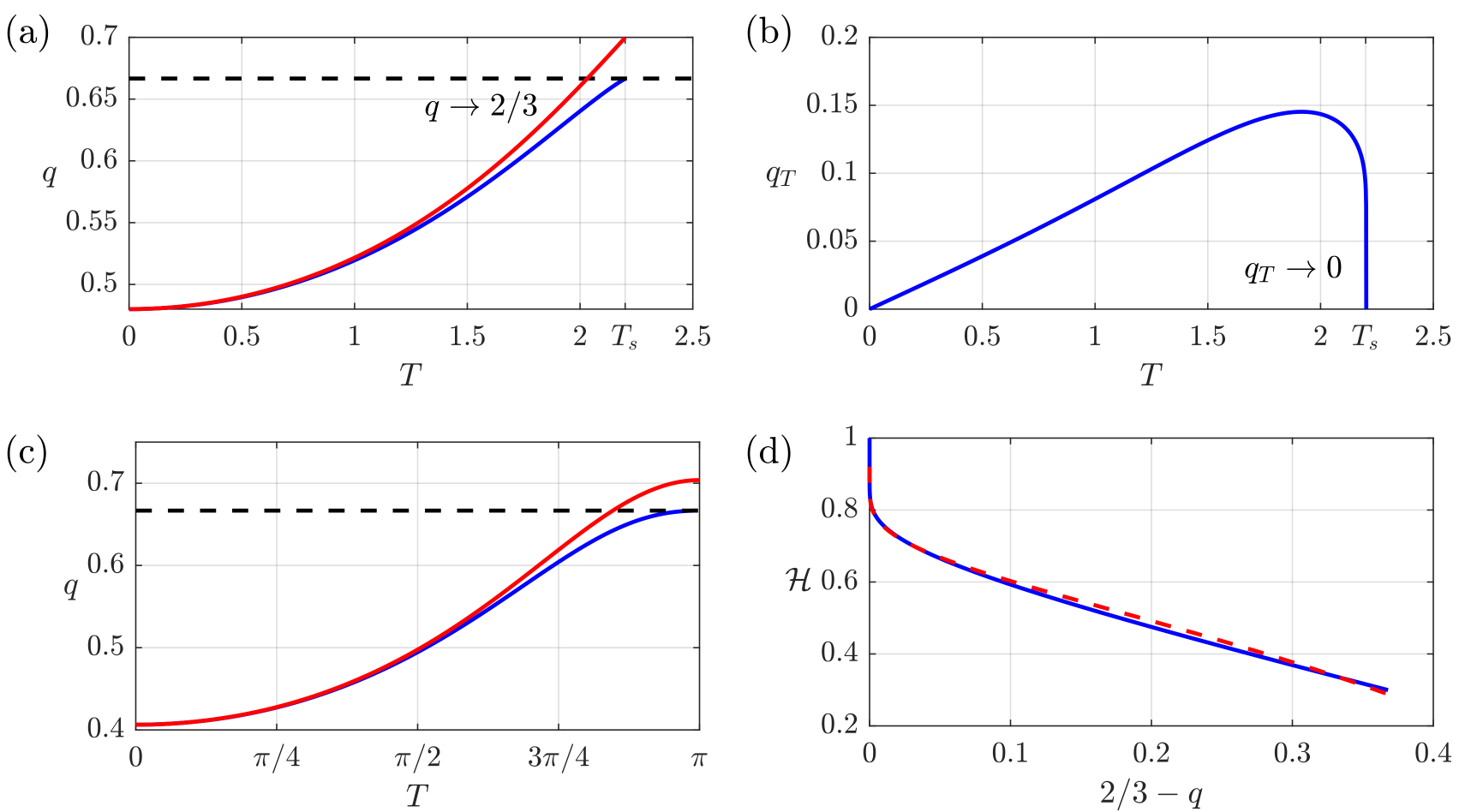}
\caption{(a, b) Numerical solution of \eqref{qeq} (solid blue curves) for $q(0)=0.48$ and $b=0.5$. The integration is stopped at $T=T_s$ where $q(T_s)=2/3$. The solid red curve shows 
the approximate $q(T)$ given in \eqref{qT_approx}. 
(c) Numerical solution of \eqref{qeq} (solid blue curve) for $b=0.5$ and 
$q(0) = q^*(0.5) \approx 0.4063$, with the approximate solution \eqref{qT_approx} 
shown with a red curve.
%$q(T)$ reaches 2/3 with zero slope at $T = \pi$. 
(d) Comparison between $\mathcal{H}$ as defined in \eqref{H0bar}, shown with a blue solid line, and 
the approximation for $\mathcal{H}$ valid as $q\to2/3$ given by \eqref{H0barapprox}, shown with a red dashed line.}
\label{fig:qTT}
\end{figure}

The nonlinear ordinary differential equation \eqref{qeq} determines $q(T)$. It has the invariant $\Omega^{1/2}\int_0^{2\pi}H_0 d\theta$, which means that fluid volume is conserved at leading order (this can be shown by differentiating \eqref{steady_mult_scales} with respect to $T$ and then by straightforward manipulations). Furthermore we can show that when $2/3-q\ll 1$ %(see Appendix B), 
\bea\label{H0barapprox}
\mathcal{H} \sim 1 + \frac{2-\left(2/3-q\right)^{1/2}}{2\log \left(2/3-q\right)}.
%\mathcal{H} \sim 1 + O\left (1/\log (2/3-q)\right).
\eea
Inserting this result into \eqref{qeq} we can then see that $q_T\to 0$ as $q\to 2/3$ and, moreover, 
$q_{TT}\to -\infty$ in the same limit. This behaviour is demonstrated numerically in figure~\ref{fig:qTT}(a,b)
for $b=0.5$ and initial condition $q(0)=0.48$. Good agreement between the approximation 
\eqref{H0barapprox}, valid when $2/3-q\ll1$, and the numerically computed $\mathcal{H}(q)$, is shown in 
figure~\ref{fig:qTT}(d).  
%{\color{red} Here to include a figure demonstrating this numerically}. 

%
%%%%%%%
%%%%%%%
%

When $q=2/3$ the leading order profile $H_0$ is described by the the heteroclinic connection along the 
separatrix that connects the saddle points at $\theta = 0$, $2\pi$ (see figure~\ref{fig:sketch}). In this case $\nu_0(\theta)\to \infty$ as $\theta\to 0$. A slight rearrangement of \eqref{flump} is
\bea \label{flump2}
\int_\pi^\theta R(\xi)\,d\xi 
= \Omega^{1/2} \nu_0(\theta) \left( q_T +
\frac{\Omega_T}{\Omega}\left[\frac{3}{2} q - 1\right]\right)
+
\frac{\Omega_T}{\Omega^{1/2}} \int_\pi^{\theta} \frac{(1-H_{0})}{M_0}\,d\xi.
\eea
With the help of \eqref{steady_mult_scales} written for $q=2/3$ we can deduce that
\bea
\int_0^{\pi} \frac{(1-H_{0})}{M_0}\,d\theta = \frac{1}{2}\int_0^{\pi} H_0 d\theta < \infty,
\eea
the inequality being clear given that the integral represents one half of the total fluid volume on the cylinder. Setting $q=2/3$ in \eqref{flump2}, and taking the limit $\theta\to 0$, we observe that the terms in the large curved bracket on the right hand side of \eqref{flump2} must vanish. Therefore in this case (cf. \eqref{qeq} for $q<2/3$) we have
\bea\label{q2/3}
q_T = 0,
\eea
and it follows that $q=2/3$ for all $T$. Unlike \eqref{qeq}, this equation does not conserve the quantity 
$\Omega^{1/2}\int_0^{2\pi}H_0 d\theta$ since in this case the integral has a fixed value and $\Omega$ varies 
with $T$.

\begin{figure}[!t]
\centering\includegraphics[width=3.75in]{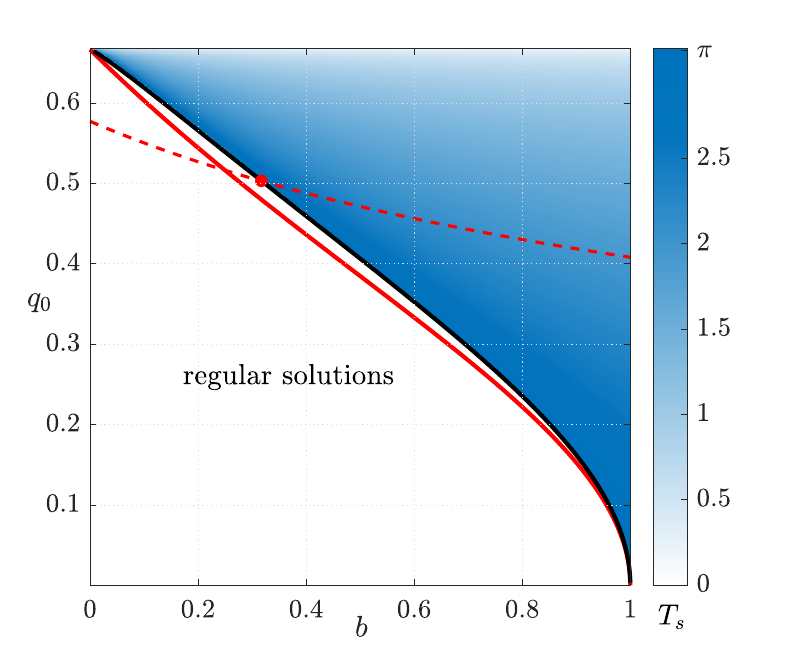}
\caption{Regularity map for $q(T)$ solving \eqref{qeq} in the limit $\sigma\to 0$. The 
black curve delineates the boundary $q^*(b)$ between regular behaviour ($q(T)$ periodic in 
$T$) and blow-up behaviour ($q(T)\to 2/3$ as $T\to T_s$). The red curve shows the 
approximation to the $q^*(b)$ boundary given in \eqref{qstareapprox}. The dashed red line 
shows $1/q_0 = \sqrt{3(1+b)}$, whose intersection with the black line at $b=0.317$, shown with a red 
marker point, determines the location of the red marker point in figure~\ref{fig:blowup_map_zooms}.} %Colormap with $T_s$.}
\label{fig:secularity_cond}
\end{figure}

Next we integrate \eqref{qeq} numerically starting from some $q(0)$. For 
$q(0)<q^*(b)$, where $q^*(b)$ is to be determined, $q$ is periodic in time $T$. For $q(0)=q^\ast(b)$, we find that $q(\pi)=2/3$ and $q_T(\pi)=0$. If 
$q(0)>q^*(b)$ then $q(T_s)=2/3$ with $q_T(T_s)>0$ at some $T_s=T_s(q_0,b)>0$. This heralds the 
breakdown of the approximation since a smooth solution of \eqref{steady_mult_scales} ceases to 
exist (see the discussion in section \ref{sec:steady}). Accordingly $q^*(b)$ descirbes a curve in the $(b,q_0)$ 
plane which divides regular solutions of \eqref{qeq} from those which blow up. 
The implicit nonlinear relation for $q^\ast(b)$,
\bea\label{qstareq}
\int_{q^\ast}^{2/3}\frac{dq}{\mathcal{H}(q)-\frac{3}{2}q} = \log\left(\frac{1-b}{1+b}\right),
\eea
follows on integrating \eqref{qeq} from $T=0$ to $T=\pi$. We solve this using Newton's method to obtain
the $q^*(b)$ curve shown with a black solid line in figure~\ref{fig:secularity_cond}.
%
%%%%%%%%%%
%%%%%%%%%%%
%

\begin{figure}[!t]
\centering\includegraphics[width=4.25in]{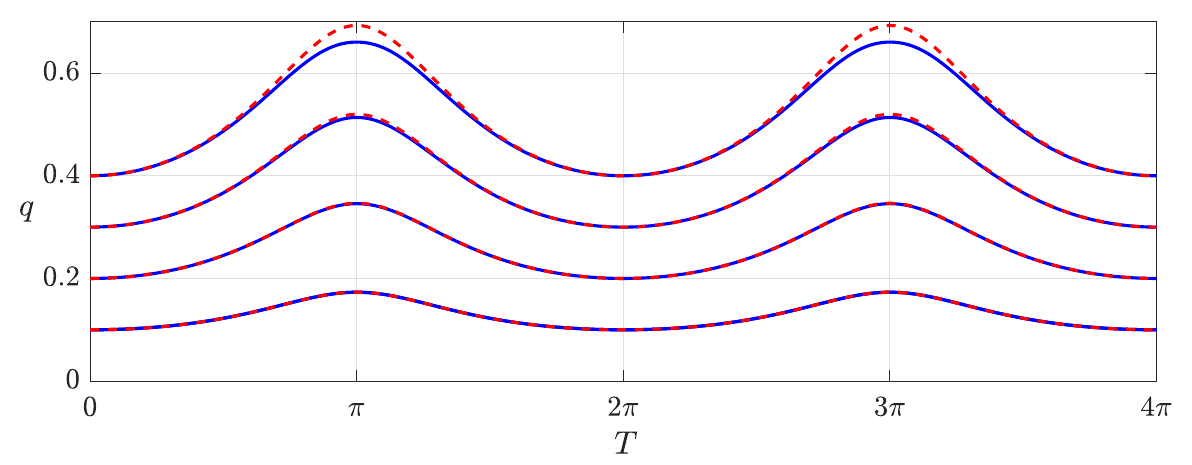}
\caption{Numerical integration (blue curves) of \eqref{qeq} for $b = 0.5$ and and $q(0) = 0.1$, $0.2$, $0.3$, $0.4$ (bottom to top). The dashed red lines show the approximation \eqref{qT_approx}.}
\label{fig:q_regular}
\end{figure}

As an alternative to integrating \eqref{qeq} numerically, a simple approximation to $q(T)$ can be found
by first noting that the one-term Taylor expansion about $q=0$, given by $\H \approx q$, 
holds good unless $q$ is very close to $2/3$ 
(this is confirmed by comparison with the numerical solution to \eqref{qstareq}).
With this approximation \eqref{qeq} simplifies to $q_T = -(\Omega_T/2\Omega)q$,
which can be integrated exactly to yield 
\bea\label{qT_approx}
q(T)=q_0\frac{\Omega_0^{1/2}}{\Omega^{1/2}},
\eea
where $q_0=q(0)$ and $\Omega_0=\Omega(0)=1+b$. This result immediately provides the 
approximation to the $q^*(b)$ curve shown with a solid red line in figure~\ref{fig:secularity_cond},
\bea\label{qstareapprox}
q^\ast(b)=\frac{2}{3}\left(\frac{1-b}{1+b}\right)^{1/2}.
\eea
In figure~\ref{fig:q_regular} we
compare the approximation \eqref{qT_approx} with the solution of the full equation \eqref{qeq}
for the case $b=0.5$ and for a number of 
different initial conditions, $q(0)$. Evidently 
the approximation performs very well with the greatest discrepancy occurring at the maxima for the 
larger values of $q(0)$. 

%%%%%%%%%%%%%%%%%%%%%%%%%%%%%%
%%%%%%%%%%%%%%%%%%%%%%%%%%%%%%
%

\begin{figure}
\centering\includegraphics[width=4.5in]{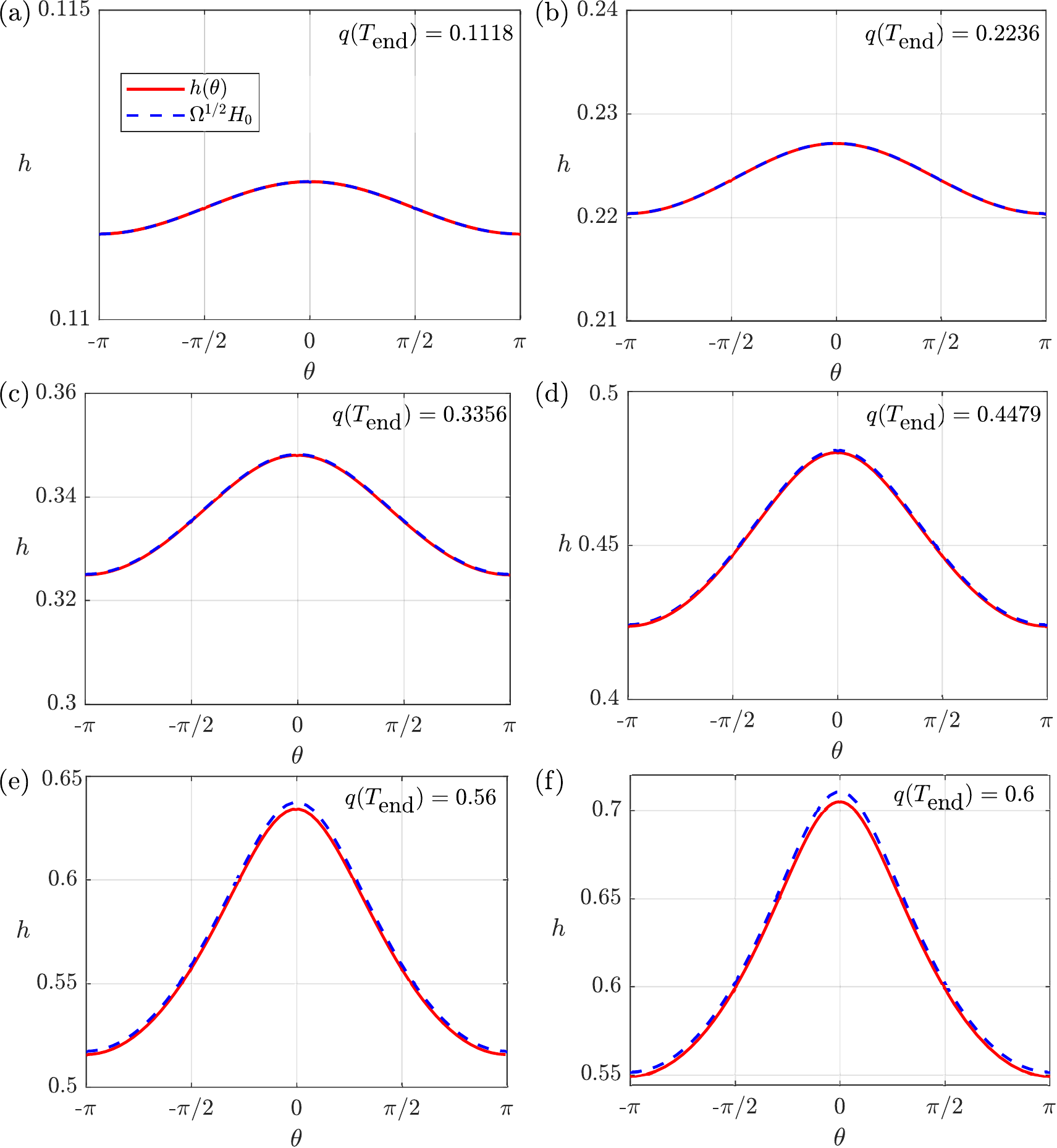}
\caption{Film profiles (solid red curves) obtained by integrating \eqref{maineq} numerically up to 
$t_{\textup{end}} = T_{\textup{end}}/\sigma = 2000/\sigma$ for the case $b=0.25$ and $\sigma = 0.01$. The 
initial condition is \eqref{ic} with $\textup{h}_0\in \mathscr{H}$ and $\mathcal{Q}_0=q_0 \Omega_0^{3/2}$ 
where $\Omega_0=\Omega(0)=1+b$ and (a) $q_0 = 0.1$, (b) $q_0 = 0.2$, (c) $q_0 = 0.3$, (d) $q_0 = 0.4$, 
(e) $q_0 = 0.5$, and (f) $q_0 = 0.5385$. For panel (f) $q_0$ is very close to the critical value $q^*(0.25) 
\approx 0.538725$. The dashed blue lines show $\Omega^{1/2}H_0$, where $H_0$ is the leading order 
solution in the small-$\sigma$ analysis of section~\ref{sec:lowfreq}, satisfying \eqref{steady_mult_scales} with 
$q(0)=q_0$ and using the approximate solution \eqref{qT_approx} for $q(T)$.}
%The function $q(T)$ was computed by integrating \eqref{qeq} numerically \crd $\leftarrow$ Antonio, I think actually this figure %uses the approximation \eqref{qT_approx}. Is that right, or did you use \eqref{qeq}?\cb. \textcolor{cyan}{AJ: You're right. The %dashed blue lines are computed using the approximation \eqref{qT_approx}}.}
\label{fig:profiles_comparison}
\end{figure}

It is instructive to compare the present small-$\sigma$ asymptotic results with numerical solutions to 
the full governing equation \eqref{maineq}. 
First, we note that for $q_0<q_*(b)$ the dependence of the solution on both $t$ and $T$ indicates that the 
small-$\sigma$ solution will in general be quasiperiodic, and this is consistent with the comment made earlier
in section~\ref{sec:dynamical_system} (see also figure~\ref{fig:quasi}).
Figure~\ref{fig:profiles_comparison} shows time snapshots of the surface profiles obtained by  
integrating the full governing equation \eqref{maineq} numerically for a variety of initial conditions 
\eqref{ic} with $\textup{h}_0\in \mathscr{H}$.
The leading order asymptotic solution, using the approximation \eqref{qT_approx} for $q(T)$, is included for 
comparison. The agreement between the two sets of results is generally excellent, but with some 
discrepancy at the surface maximum seen when $q$ gets close to $2/3$.

%
%%%%%%%%%%%%%%%%%%%%%%%%%%%%%%
%%%%%%%%%%%%%%%%%%%%%%%%%%%%%%
%
In figure~\ref{fig:Tstar_mgb}(a) we showed the numerically computed overturning time 
$t^*$ for the case $b=0.5$ and $q_0 = 0.5$. These parameter values correspond to a point in the white 
regular region in figure~\ref{fig:secularity_cond}. We can match the expansion \eqref{smallsigexp} to the 
initial condition used in figure~\ref{fig:Tstar_mgb}(a) by taking $H_0(\th,0,0)=\textup{h}_0$, and $H_n(\th,0,0)=0$ for $n\geq 1$ (the latter can be effected by choosing the arbitrary functions $U_n$ in 
\eqref{small_sig_gen_sol} appropriately). Accordingly the asymptotics predict no overturning in the limit $\sigma\to 0$ and this is in line with the very rapid growth of $t^*$ observed in 
figure~\ref{fig:Tstar_mgb}(a) in this limit.
In contrast figure~\ref{fig:Tstar_mgb}(b) shows the overturning time for the same value of $b$ but with the initial profile solving $\textup{h}_0 - (1/3)\textup{h}_0^3\cos\theta = 1/2$. Matching the asymptotic 
expansion \eqref{smallsigexp} to this initial condition, at leading order we find that 
\eqref{steady_mult_scales} is not satisfied at $T=0$ and, consequently, the solution of 
the leading order equation \eqref{smallsigH0} will develop a slope singularity at a finite time, as was 
discussed in section~\ref{sec:steady}. The predicted overturning time $t^*=242$ obtained by integrating 
\eqref{smallsigH0} with $\Omega$ fixed at its initial value $\Omega(0)=1+b$, is shown with a dashed line in 
figure~\ref{fig:Tstar_mgb}(b).
%
% The value t^*=242 comes from running MGB's code manyth_ode45_varysig_2.m in 
% Matlab/characteristics/Origin_new/z-results/asymptotic_compare
%

The blow-up that occurs at $T_s=T_s(q_0,b)$ in the blue region in figure~\ref{fig:secularity_cond} (i.e. 
for $q^*(b)<q_0<2/3$) can be interpreted as the overturning of the film profile in the full problem. At $T=T_s$ we 
expect the expansion \eqref{smallsigexp} to disorder, yielding the approximation to the overturning time for 
the full problem $t^*= T_s(q_0,b)\sigma^{-1}$ valid for $\sigma \ll1$. In figure~\ref{fig:Tstar_mgb}(c) we 
showed the overturning time for the case $b=0.5$, $q_0=\sqrt{2}/3=0.471$ corresponding to a point in the 
blow-up region in figure~\ref{fig:secularity_cond}. With this value of $q_0$, integrating \eqref{qeq} we compute 
$T_s=2.270$ and hence the prediction $t^* = 2.270\sigma^{-1}$. This is shown with a red dashed line in 
figure~\ref{fig:Tstar_mgb}(c). At sufficiently small $\sigma$ It is in excellent agreement with the numerically 
computed overturning time for the full governing equation. 

Finally, we note that the present results are consistent with the blow-up map shown earlier in 
figure~\ref{fig:blowup_map_zooms}. For this figure, trajectories solving \eqref{hamilform} were 
computed with the initial condition \eqref{blowup_ic} and $\textup{h}^*=1/\sqrt{3}$. This means 
that, in the context of the present small $\sigma$ analysis, the relevant trajectories start on the 
$(\theta,H_0)$ curve with $q_0=1/\sqrt{3(1+b)}$. This curves is shown with a red dashed line in figure ~\ref{fig:secularity_cond}, and where it intersects the 
black curve representing $q^*(b)$ gives the threshold $b$ value for 
small-$\sigma$ blow-up in figure~\ref{fig:blowup_map_zooms}. The intersection is found to occur at $b=0.317$. This  is shown with a red 
marker in figure~\ref{fig:blowup_map_zooms}, where it can be seen to be in good agreement with the boundary curve between regular and blow-up behaviour at small $\sigma$.

\subsubsection{Single shock solution for $q = 2/3$}\label{sec:shocks}

Under the small-$\sigma$ assumption, according to \eqref{q2/3} if $q(0)=2/3$ then $q=2/3$ for all $T$. %The corresponding film profile has a corner at its maximum point.
If we take the surface profile for $q=2/3$ to be $H_0$ satisfying 
\eqref{steady_mult_scales} and given by
the solid gold separatrix curve in figure~\ref{fig:pplane}, a contradiction arises since conservation of mass at leading order requires $\Omega^{1/2}\int_0^{2\pi}H_0 d\theta$ and $\Omega$ is $T$-dependent. 
%In this case the surface profile has a corner at its maximum point corresponding to the location of the saddle.
We can circumvent the mass conservation issue by following Johnson \cite{johnson1988steady} to introduce a shock at a location 
$\theta^{\textup{s}}(T)>0$. The film profile follows the solid separatrix curve in figure~\ref{fig:pplane}
over $-\pi\leq \theta\leq \theta^{\textup{s}}$ (proceeding 
smoothly through the saddle at $\theta=0$ onto the dashed part of the separatrix curve where $H_0>1$) and then, at 
$\theta=\theta^{\textup{s}}$, it drops vertically down to the solid part of the separatrix below and continues along this up to $\theta=\pi$. 
%The shock location $\theta^{\textup{s}}(T)$ must be such that so as to conserve mass at leading order.

A shock solution allows the cylinder to support a larger liquid volume than a smooth solution. The liquid 
volume increases monotonically as $\theta_s$ varies from $0$ for which $V^*/2\pi \approx0.70708140$, up to $\theta_s=\pi/2$ for which value the film has infinite thickness at the shock but the finite volume
\bea\label{maxVolshock}
V_{\text{max}} = 6\Omega^{1/2}\sum_{n=1}^3\int_{a_n}^{b_n} \frac{dH}{(H+2)^{1/2}(H^3+3H-2)^{1/2}},
\eea
where $(a_1,b_1) = (H_m,1)$, $(a_2,b_2) = (1,\infty)$, $(a_3,b_3) = (H_m,2/3)$. Here $H_m$ is the 
minimum of $H$ on the separatrix which occurs at $\theta=\pm\pi$. 
We calculate $V_{\text{max}}/(2\pi \Omega^{1/2}) \approx 1.102317$, which agrees with the value quoted by Villegas-Diaz {\it 
et al.} \cite{villegas2005shocks}.

Suppose that we start at $T=0$ with a leading order profile $H_0$ featuring a shock located at 
$\theta=\theta^{\textup{s}}(0)$ with $0<\theta^{\textup{s}}(0)\leq \pi/2$. The fluid volume $V(T) = \int_0^{2\pi}H_0 d\theta$. As time increases the shock location $\theta^s(T)$ must adjust to ensure mass conservation, i.e. such that $\Omega^{1/2} V(T) = (1+b)^{1/2}V(0)$. The requirement that $|\theta^{\textup{s}}|<\pi/2$ imposes the constraint
\bea
\left(\frac{1+b}{1-b}\right)^{1/2} \frac{V(0)}{V_{\text{max}}} \leq 1.
\eea
This places an upper limit on the modulation amplitude, $b$, for a given 
initial volume.

\subsubsection{Double shock solution for $q < 2/3$}

Up to this point the discussion for $q(T)<2/3$ has assumed that the leading order term $H_0$ is such 
that $\Omega^{1/2}H_0\in \mathscr{H}$ at $t=T=0$ with $H_0$ a smooth function (i.e. no shocks).
Johnson \cite{johnson1988steady} demonstrated that one can construct double-shock solutions for which the film profile exhibits a bulge in fluid thickness enclosed by the two shocks. Referring to 
the phase portrait in figure~\ref{fig:pplane}, we see that a double-shock solution can be put together by 
following one of the blue trajectories at the bottom of the figure for a chosen $Q$, and then, somewhere in 
the interval $-\pi/2<\theta<\pi/2$, jumping up to and then back down from the equivalent $Q$ curve at the 
top of the figure, and then continuing along the original trajectory to complete the profile. It is not necessary that the jumps be 
located symmetrically about $\theta=0$.

A case of particular interest is that of zero flux, $Q = 0$, indicated by the broken blue line in figure~\ref{fig:pplane}. For $Q=0$, 
and in the absence of shocks, the only static solution is a film of zero thickness. For the double-shock solution a solitary mass 
of fluid with compact support is carried around by the cylinder, with the fluid inside recirculating (Johnson 
\cite{johnson1988steady} describes how to extend the thin-film model to include next order terms and 
allow for a boundary layer around each shock to facilitate this recirculation).
The maximum scaled volume that can be supported in this case can be expressed by
\bea\label{volumeQ0}
\frac{V_{\text{max}}^0}{2\pi} 
%= \frac{\sqrt{3}}{2\pi}\int_{-\pi/2}^{\pi/2} \frac{d\th}{\sqrt{\cos\th}} 
= \frac{\sqrt{3}}{\pi}\int_0^{\pi/2} \frac{d\th}{\sqrt{\cos\th}}
= \frac{2\sqrt{3}}{\pi}F\left(\frac{\pi}{4}; \sqrt{2}\right)\approx 1.44562,
\eea
%The last RHS uses the property of $1/\sqrt{\cos\th}$ as an even periodic function around $\th=0$.
where $F$ is the incomplete elliptic integral of the first kind (e.g. Olver {\it et al.} \cite{Olver.etal.2010}).

Assume for our small-$\sigma$ problem that a double shock is present in the leading order 
profile $H_0$. Let the jumps be located at $\theta = \theta_L<0$ and $\theta_R>0$ with 
$|\theta_{L,R}|\leq \pi/2$. Since $\Omega^{1/2} \int_0^{2\pi}H_0 d\theta$ is automatically conserved as $q$ evolves 
according to \eqref{qeq}, the shock locations $\theta_{L,R}$ do not depend on $T$ and so remain fixed.
If $q(0)<q^*(b)$ then $q(T)$ is $T$-periodic and will oscillate in the range $(0,2/3)$ with the double 
shock profile adjusting accordingly. If $q(0)\geq q^*(b)$ then $q(T)\to 2/3$ as $T\to T_1$, for some 
finite $T_1$, with $q_T(T_1)>0$; in this case, we jump to the single shock solution described in the previous subsection.

%%%%%%%%%%%%%

\section{Conclusion}\label{sec:conclusion}

We have examined the dynamics of a thin viscous liquid film flowing on the outside of a horizontal cylinder 
which is rotating about its axis with an angular velocity that includes a steady part and a time-periodic part. 
Surface tension has been neglected. If the time-periodic component of the angular velocity is removed, the 
problem reduces to that studied by Moffatt \cite{moffatt1977behaviour} and Pukhnachev 
\cite{pukhnachev1977motion}, and subsequent authors.

Our brief review of the constant rotation rate Moffatt-Pukhnachev problem included a novel perspective on 
describing steady solutions using a phase-plane analysis. This perspective makes plain visually the steady 
solution space, including both smooth solutions and those exhibiting shocks. It also makes clear graphically 
the threshold on the rotation rate for a steady solution to exist, and the onset of dripping that is expected 
if this threshold is exceeded. A separatrix in the relevant phase portrait delineates a boundary representing 
the extreme continuous steady solution which supports the maximum fluid volume (although larger volumes are 
possible for solutions with shocks).

The introduction of a time-periodic part into the angular velocity fundamentally alters the behaviour of the 
system. In this case it appears that in general it is not possible to choose an initial condition such that the slope of the 
film profile remains finite for all time. Rather it seems that the dynamics inevitably progress to a point where 
the film starts to overturn and a shock is formed. Numerical investigation of the characteristic dynamical 
system revealed a highly intricate and complex blow-up map showing the tendency toward film overturning or 
blow-up depending on the amplitude and frequency of the angular velocity modulation.
Notably, sharp protrusion-like structures in these maps appear near to the base frequency $\omega^*$ of the 
Moffatt-Puchnachev system and, seemingly, at rational multiples thereof. 

To develop a deeper understanding of the dynamics we also performed asymptotic analyses in the limits of high 
and low forcing frequency. In latter case a multiple-scales analysis suggested that the flow is time-periodic if  
the initial condition is chosen sufficiently carefully. For a general initial condition the film profile will overturn; 
but this overturning can be considerably delayed by choosing the initial condition to coincide with a 
Moffatt-Puchnachev steady profile. A multiple-scales approach was also used in the low-frequency limit. In 
this case a secularity condition was derived, which yielded a threshold modulation amplitude beyond which a 
singularity is reached at some time. This singularity, which heralds 
the breakdown of the multiple-scales expansion, was interpreted as indicating the blow-up of 
the solution to the full problem, and confirmation of this point was provided through comparison with 
numerical solutions of the full problem. Regular, bounded solutions exist below the threshold amplitude, and 
it was shown that these correspond to quasiperiodic solutions of the full problem. Single-shock and double-
shock solutions for which the film jumps instantaneously in height, were also discussed. In both cases the 
abrupt change in height can be smoothed by introducing boundary layers as discussed by 
Johnson \cite{johnson1988steady}.

In summary our results have revealed a complex interplay between the amplitude and frequency of the 
torsional movement of the cylinder and the stability and regularity of the liquid film. 
%Our numerical 
%calculations have highlighted apparent \cbl {\it resonant-like behaviours at frequencies related to the intrinsic 
%dynamics of the steady flow}\cb, and our asymptotic analyses at the high- and low-frequency limits have 
%provided valuable insight into the conditions governing the periodicity/quasiperiodicity and potential blow-up 
%of the film model. The dynamics depend sensitively on the frequency of the oscillations. At high frequency the 
%oscillations tend to regularise the flow, favouring periodic and bounded solutions, provided that the inital 
%condition is chosen carefully. At low frequency, the oscillations can induce blow-up if the oscillation amplitude is sufficiently large. \crd repetition?\cb
As in Moffatt's original work we have herein neglected the effect of surface tension. The influence of this force 
on the dynamics of the film in the presence of a modulated rotation rate is the subject of our ongoing research.

\section*{Appendix}\label{app:Mpositive}
We demonstrate that $M=\Omega - h_s^2 \cos\theta$ is positive for all $\theta\in [0,2\pi)$, where $h_s(\theta)$ satisfies
\bea \label{app_hs}
\Omega h_s - \frac{1}{3}h_s^3 \cos\theta = Q, 
\eea
for constant $\Omega$, provided that $Q<(2/3)\Omega^{3/2}$.
Writing $h_s = \Omega^{1/2}H$, $Q=\Omega^{3/2}q$, and manipulating \eqref{app_hs}, we have
\bea
M = \frac{2\Omega}{H}\left(\frac{3}{2}q - H\right),
\eea
where $q<2/3$.
The case $\cos\theta=0$ is clear since then $H=q$. Assuming $\cos\theta > 0$ and writing $H=(\cos\theta)^{-1/2} R$, \eqref{app_hs} becomes
\bea \label{gprob}
g(R) = \lambda,
\eea
where $\lambda=(\cos\theta)^{1/2}q<2/3$ and $g(x)\equiv x - x^3/3$. The cubic equation \eqref{gprob} has three real roots, and it's clear that the root of interest is that for which $0<R<1$ (refer to Figure~\ref{fig:pplane}; the other positive root has $R>1$ and will produce blow-up). Evidently $g(x)>2x/3$ for $0<x<1$ since over this range $g(x)-2x/3 = x(1-x^2)/3>0$. It follows that $\lambda>2R/3$ and, consequently, $H<3q/2$ so that $M>0$. A similar, but slightly simpler, argument applies in the case when $\cos\theta<0$. 

\section*{Acknowledgments}
This work was partially supported by the research project, PID2020-115961RB-C31 financed by MCIN/AEI/10.13039/501100011033. AJBL would like to thank the Spanish Ministry of Science, Innovation and Universities  for the financial support provided by the Fellowship PRE2021-099112 that allowed his research stay in the School of Mathematics at the University of East Anglia.

\bibliographystyle{unsrt} % estilo de la biblio
\bibliography{bibliography} % nombre de tu archivo .bib (sin .bib)

%\begin{thebibliography}{39}
%\providecommand{\natexlab}[1]{#1}
%\providecommand{\url}[1]{\texttt{#1}}
%\expandafter\ifx\csname urlstyle\endcsname\relax
%  \providecommand{\doi}[1]{doi: #1}\else
%  \providecommand{\doi}{doi: \begingroup \urlstyle{rm}\Url}\fi

%\bibitem[Linninger et al.(2016)]{Linninger2016}
%A.A. Linninger, K. Tangen, C.Y. Hsu, and D. Frim. 
%\newblock Cerebrospinal fluid mechanics and its coupling to cerebrovascular dynamic
%\newblock \emph{Annu. Rev. Fluid Mech.}, 48\penalty0 (1): \penalty0 219–-257, 2016.

%\end{thebibliography}

\end{document}